\documentclass[aps,prd,twocolumn]{revtex4}

\usepackage{graphicx}
\usepackage{amssymb}

\newcommand \tg{\tilde\gamma}
\newcommand \tG{\tilde\Gamma}
\newcommand \tA{\tilde A}
\newcommand \p{\partial}

\newcommand \g{\gamma}
\newcommand \br{\nonumber \\ &&}

\begin{document}

\title{Well-posedness of formulations of the Einstein equations with
  dynamical lapse and shift conditions}

\author{Carsten Gundlach}
\affiliation{School of Mathematics, University of Southampton,
         Southampton SO17 1BJ, UK}

\author{Jos\'e M. Mart\'\i n-Garc\'\i a}
\affiliation{Instituto de Estructura de la Materia, Centro de F\'\i sica
Miguel A. Catal\'an, C.S.I.C., Serrano 123, 28006 Madrid, Spain}


\begin{abstract}

We prove that when the equations are restricted to the principal
part the standard version of the BSSN formulation of the Einstein
equations is equivalent to the NOR formulation for any gauge, and that
the KST formulation is equivalent to NOR for a variety of gauges. We
review a family of elliptic gauge conditions and the implicit
parabolic and hyperbolic drivers that can be derived from them, and
show how to make them symmetry-seeking. We investigate the
hyperbolicity of ADM, NOR and BSSN with implicit hyperbolic lapse and
shift drivers. We show that BSSN with the coordinate drivers used in
recent ``moving puncture'' binary black hole evolutions is ill-posed
at large shifts, and suggest how to make it strongly hyperbolic for
arbitrary shifts. For ADM, NOR and BSSN with elliptic and parabolic
gauge conditions, which cannot be hyperbolic, we investigate a
necessary condition for well-posedness { of the initial-value problem.}

\end{abstract}


\date{\today}

\maketitle



\section{Introduction}


Numerical solutions of the Einstein equations are often obtained by
specifying initial data on a hypersurface and evolving in time. The
Einstein equations then split into evolution equations which contain
time derivatives and constraints which do not. A choice of variables
and of evolution equations for them is called a formulation of the
Einstein equations. At the same time, some metric components are not
determined by the evolution equations or constraints and must be
determined otherwise to obtain a closed system. This is called a
choice of gauge.

A standard formulation is the ADM one, whose variables are the metric
$\g_{ij}$ and extrinsic curvature $K_{ij}$ of a spacelike
hypersurface. The ADM evolution equations for these two tensors are
first order in time and second order in space. The ADM constraint
equations are second-order in space. The $\g_{ij}$ represent 6 out of
10 components of the metric $g_{\mu\nu}$ of the spacetime. The other 4
components can be parameterised by a scalar $\alpha$, the lapse, and a
3-vector $\beta^i$, the shift. The lapse and shift do not appear in
the ADM constraints. They appear in the evolution equations, but
without time derivatives. They must therefore be determined by a gauge
choice.

One distinguishes fixed gauge choices, in which the lapse (or the
densitised lapse) and shift are given a priori as functions of the
coordinates $x^\mu$, and live gauge conditions. The latter can be
subdivided into algebraic ones where the lapse and shift are
algebraically related to the dynamical variables, differential
(typically elliptic) constraints on the lapse and shift, and evolution
equations for them.

In order for numerical solutions to converge to the continuum
solution, given that discretisation error in the evolution equations
generates constraint violations, the continuum time evolution problem
must be well-posed for arbitrary initial data that do not obey the
constraints. In this paper we investigate the well-posedness of a
number of live gauge conditions combined with the ADM formulation and
three formulations that are derived from it, the NOR, BSSN and KST
formulations.

In Sect.~\ref{section:formulations} we review the formulations while
leaving the gauge still undetermined. We show that, roughly
speaking, the NOR formulation is equivalent in the principal part to
the BSSN formulation and the KST formulation. For the KST formulation
this will require shedding three variables that evolve trivially.

In Sect.~\ref{section:gauges} we review live gauge conditions. Our
guiding principle is that the gauge should be
``symmetry-seeking''. Specifying the derivatives of the lapse and
shift along the normal to the time slices makes it easier to
establish well-posedness analysis but specifying their time
derivative makes it easier to show a gauge is symmetry-seeking. We
show that the two time-derivatives are equivalent in some cases.

In Sections~\ref{section:hyperbolicity}-\ref{section:modeanalysis} we
combine formulations and gauges, and investigate the well-posedness of
the resulting closed systems. What we can prove depends on the choice
of gauge. In Sect.~\ref{section:hyperbolicity} and
\ref{section:puncturegauge} we investigate gauges in which the whole
system is strongly or symmetric hyperbolic. This gives us both
necessary and sufficient conditions for well-posedness.  In
Sect.~\ref{section:modeanalysis} we use mode analysis to find a
necessary condition for the well-posedness of the Cauchy problem of
the remaining, non-hyperbolic,
systems. Sect.~\ref{section:conclusions} contains our conclusions.


\section{Equivalence of formulations of the Einstein equations}
\label{section:formulations}


\subsection{Definitions}


{ Well-posedness of an initial value problem implies that
an estimate
\begin{equation}
\label{estimate}
||\delta u(\cdot, t)|| \le F(t) ||\delta u(\cdot, 0)||
\end{equation}
exists, where $u(x,t)$ is the solution, $u(x,0)$ the initial data,
$\delta$ denotes a linear perturbation and $||\cdot||$ stands for
appropriate norms (which may involve spatial derivatives), and where
$F(t)$ is independent of the initial and boundary data. This means
that the solution depends continuously on the initial and boundary
data.
Necessary or sufficient conditions for the well-posedness of the
initial-value problem, such as strong hyperbolicity, can usually be
determined from the principal part of the evolution equations.}

In this Section we discuss formulations of the 3+1 Einstein equations
{ derived from the ADM formulation} without yet specifying the lapse
and shift. { These are systems of evolution equations which are
first-order in time in all evolved variables, but second-order in space
in some variables $v$ and $z$ and first-order in other variables $w$.}
We shall therefore be looking at quasilinear systems of equations
whose principal part is
\begin{eqnarray}
\label{vdot}
\dot v &\simeq& \p v + w + \p z, \\
\label{wdot}
\dot w &\simeq& \p\p v + \p w + \p\p z
\end{eqnarray}
where $\simeq$ indicates equality in the principal part, a dot denotes
$\p_t\equiv \p/\p t$, and $\p$ stands for spatial derivatives. We will
close this system later by providing evolution or constraint equations
for the (gauge) variables $z$.

{ We shall show that for certain parameter choices and in certain
gauges these formulations are equivalent, roughly in the sense that
there is a bijection between their variables and their principal parts,
and that the initial-value problem for one system is well-posed if and
only if it is well-posed for the other. The detailed results will be
stated in a Lemma at the end of each subsection.}


\subsection{ADM}


We write the spacetime metric in the well-known 3+1 form
\begin{equation}
ds^2=-\alpha^2\,dt^2+\g_{ij}(dx^i+\beta^i\,dt)(dx^j+\beta^j\,dt).
\end{equation}
The ADM formulation of the Einstein equations { in the form given
by York} \cite{York} is 
\begin{eqnarray}
\label{ADM1}
\partial_t \gamma_{ij} &=& {\cal L}_\beta \gamma_{ij} -2\alpha K_{ij},
\\ 
\nonumber \partial_t K_{ij} &=& {\cal L}_\beta K_{ij} -D_iD_j\alpha \\
&&+\alpha\left(R_{ij}-2K_{il}{K^l}_j+KK_{ij}\right), \label{ADM2}\\
\label{Hdef}
H&\equiv &R-K_{ij}K^{ij}+K^2\doteq 0, \\
\label{Midef}
M_i&\equiv &D_j{K^j}_i-D_iK\doteq 0,
\end{eqnarray}
where $D_i$ is the 3-dimensional covariant derivative compatible with
$\g_{ij}$, $\cal L_\beta$ is the Lie derivative along $\beta^i$, and
$R_{ij}$ is the Ricci tensor of $\g_{ij}$. Indices are moved
implicitly with $\g_{ij}$ and $\g^{ij}$ throughout this paper. (Note
the distinction between $\equiv$, which denotes the definition of a
shorthand notation, and $\doteq$, which denotes a constraint or,
below, the definition of an auxiliary variable.) For convenience of
notation we define the shorthands
\begin{equation}
d_i\equiv \g^{jk} \g_{ij,k} , \quad
t_i\equiv \g^{jk} \g_{jk,i} ,
\end{equation}
where a comma denotes a partial derivative. With these the
principal part (second derivatives of $\g_{ij}$) of $R_{ij}$ is
\begin{equation}
R_{ij}\simeq -{1\over 2}\left({\g_{ij,k}}^{,k}+t_{(i,j)}\right)+d_{(i,j)}.
\end{equation}
The ADM evolution equations are first-order in time in the variables
$v=\g_{ij}$ and $w=K_{ij}$. The highest spatial derivatives in the ADM
constraints and the ADM evolution equations are second spatial
derivatives of $\g_{ij}$ and $\alpha$, and first derivatives of
$K_{ij}$ and $\beta^i$. The principal part of the Hamiltonian and
momentum constraints are
\begin{eqnarray}
H&\simeq&  {d_i}^{,i}-{t_i}^{,i}, \\
M_i&\simeq& {K_{ij}}^{,j}-K_{,i}.
\end{eqnarray}
{ Throughout this Article we consider the vacuum Einstein
equations. The generalisation to matter is trivial in the typical
case, for example perfect fluids or electromagnetism, where the
Einstein equations and the matter equations couple only through lower
order terms.}


\subsection{NOR}


The NOR formulation \cite{NOR} is obtained from the ADM formulation by
introducing the auxiliary variables
\begin{equation}
\label{fdef}
f_i\doteq d_i-{\rho\over 2} t_i,
\end{equation}
where $\rho$ is a constant parameter. This gives rise to the
auxiliary constraints
\begin{equation}
\label{Gdef}
G_i\equiv
f_i-d_i+{\rho\over 2}t_i \doteq 0.
\end{equation}
The full NOR
evolution equations are defined as
\begin{eqnarray}
\p_t \g_{ij} &=& {\rm ADM}, \label{NOR1} \\
\p_t K_{ij} &=& {\rm ADM} + a \alpha G_{(i,j)},
+\alpha \g_{ij} \left(cH+d{G_k}^{,k}\right) , \label{NOR2} \\
\p_t f_i &=& \p_t \left(d_i-{\rho\over 2}t_i\right)_{\rm ADM} +
{\cal L}_\beta G_i + 2b\alpha M_i, \label{NOR3}
\end{eqnarray}
{ where ADM in equations (\ref{NOR1},\ref{NOR2}) represents the
right hand sides of (\ref{ADM1},\ref{ADM2}) respectively, and the
subscript ADM in (\ref{NOR3}) means that the time derivative of $d_i$
and $t_i$ must be replaced by their expressions obtained from equation
(\ref{ADM1}).}
$a$, $b$, $c$ and $d$
are constant parameters to be fixed later. (We have replaced the
parameter $a'$ of \cite{bssn2} by $d\equiv ca'$.)

In the following we use the shorthand
\begin{eqnarray}
\label{defD0}
\partial_0&\equiv& \alpha^{-1}(\partial_t -\beta^k \partial_k),
\end{eqnarray}
to hide the transport terms in the principal part of the evolution
equations. We also densitise the lapse as
\begin{equation}
\alpha\equiv|\g|^{\sigma/2}Q, 
\end{equation}
where $Q$ rather than $\alpha$ is now considered as the dynamical
variable. (Later we will often set $\sigma=0$, so that
$Q=\alpha$. However, $\sigma>0$ is necessary to obtain a hyperbolic
formulation with fixed lapse and shift.)
The principal part of the NOR evolution equations is then
\begin{eqnarray}
\label{gdot}
\partial_0 \g_{ij}&=&-2K_{ij}
+2\alpha^{-1}{\beta^k}_{,(i}\g_{j)k} \\
\label{Kijdot}
\partial_0 K_{ij}&\simeq& -(\ln Q)_{,ij}
+a f_{(i,j)}-{1\over 2}{\g_{ij,k}}^{,k}
\nonumber \\ &&
+(1-a)d_{(i,j)}
+{1\over 2}(a\rho-1-\sigma)t_{(i,j)} 
\br 
+\g_{ij}
\left(cH+d{G_k}^{,k}\right), \\
\label{fdot}
\partial_0 f_i&\simeq&2(b-1){K_{ij}}^{,j}+(\rho-2b)K_{,i} \br 
+\alpha^{-1}\g_{ik}{{\beta^k}_{,j}}^{,j}+(1-\rho)\alpha^{-1}{\beta^j}_{,ij}.
\end{eqnarray}
In contrast to the ADM evolution equations, the NOR evolution
equations contain second spatial derivatives of the shift in the
evolution equation for $f^i$. We define $v=\g_{ij}$, $z=(\ln Q,\beta^i)$
and $w=(K_{ij},f_i)$. With $a=b=c=d=0$, the evolution of $f_i$ can be
ignored, and ADM is recovered as a limiting case of NOR.

The Hamiltonian constraint of the NOR system is { defined only up
to use of the constraint $G_i$. For definiteness, we denote by $H$ the
ADM Hamiltonian defined in (\ref{Hdef}) in terms of $\g_{ij}$ and
$K_{ij}$ but not $f_i$.} The evolution equations imply a closed
constraint evolution system whose principal part is
\begin{eqnarray}
\partial_0 H &\simeq& -2M{_i}^{,i}, \label{NORc1} \\
\partial_0 M_i &\simeq&-\left({1\over 2}+2c\right)H_{,i} 
\nonumber \label{NORc2} \\ && 
+{a\over 2} {G_{i,j}}^{,j}-\left({a\over 2}+2d\right){G_{j,i}}^{,j},
\\
\label{Gdot}
\partial_0 G_i&\simeq & 2 b M_i .
\end{eqnarray}
In particular, the ADM constraint evolution is given by
(\ref{NORc1},\ref{NORc2}) with $a=b=c=d=0$. { (The variable $f_i$
  then decouples from $\g_{ij}$ and $K_{ij}$ and can be neglected).}


\subsection{BSSN}


The BSSN formulation (see for example \cite{Yo2}) is obtained from the
ADM formulation by introducing the new variables
\begin{eqnarray} 
\label{tgdef}
\tg_{ij}&\doteq&(\det\gamma)^{-1/3}\gamma_{ij}, \\
\tG^i&\doteq&\tg^{ij}\tg^{kl}\tg_{jk,l}, \\
\label{phidef}
\phi&\equiv&{1\over12}\ln\det\gamma, \quad K\equiv \g^{ij} K_{ij}\\
\label{tAdef}
\tA_{ij}&\doteq&(\det\gamma)^{-1/3}\left(K_{ij}-{1\over
3}\gamma_{ij}K\right).
\end{eqnarray}
With a densitised lapse we then have $\ln\alpha=\ln Q+6\sigma\phi$.
The definition of
the $\tG^i$ gives rise to the differential constraint
\begin{equation}
\label{BSSNGdef}
G_i\equiv \tg_{ij}\tG^j-\tg^{jk}\tg_{ij,k} \doteq 0.
\end{equation}
The definition of $\tA_{ij}$ gives rise to the
algebraic constraint
\begin{equation}
{T}\equiv \tg^{ij}\tA_{ij}\doteq0,
\end{equation}
and from the definition of $\tg_{ij}$ we have the algebraic constraint
\begin{equation}
D\equiv \ln \det \tg\doteq 0.
\end{equation}
(There are no definition constraints associated with the new
variables $\phi$ and $K$ because $\g_{ij}$ and $K_{ij}$ are no longer
variables.)  Here we do not define the full BSSN evolution equations
but only their principal part. It is
\begin{eqnarray}
\partial_0 \phi &\simeq& -{1\over 6} K+ \frac{1}{6\alpha}\beta^i{}_{,i},
\\
\partial_0 \tg_{ij} &\simeq& -2\tA_{ij}
+2\alpha^{-1}\left({\beta^k}_{(,i}\tg_{j)k}-{1\over 3}
\tg_{ij}{\beta^k}_{,k}\right), \\
\label{Kdot}
\partial_0 K &\simeq& -[(\ln Q)_{,ij}+6\sigma\phi_{,ij}]e^{-4\phi}\tg^{ij}\\
\label{Aijdot}
\nonumber \partial_0 \tA_{ij} &\simeq&
e^{-4\phi}\left[
-(\ln Q)_{,ij}
- \frac{1}{2} \tg^{mn} \tg_{ij,mn} 
- 2 (1+3\sigma) \phi_{,ij}
\right. \\ && \left.
+ a \tg_{k(i} \tG^k{}_{,j)}
+ (1-a) \tg^{kl} \tg_{k(i,j)l}
\right]^{\rm TF}, \\
\label{tGdot}
\partial_0 \tG^i &\simeq& 
2(b-1)\tg^{ij}\tg^{kl}\tA_{jk,l}-{4\over 3}b\tg^{ij}K_{,j}
\br
+\alpha^{-1}\tg^{jk}{\beta^i}_{,jk}
+\frac{1}{3}\alpha^{-1}\tg^{ik}{\beta^j}_{,jk},
\end{eqnarray}
where TF indicates the tracefree part. (The entire right-hand side of
(\ref{Aijdot}) is made tracefree, not only the principal part.) The
parameters $a$ and $b$ have been introduced for comparison with
NOR. In standard BSSN, $a=b=1$. 

For definiteness, we define the Hamiltonian and momentum constraint
for the BSSN system by substituting the inverse of the definitions
(\ref{tgdef}) and (\ref{tAdef}) into the ADM definitions of these
constraints. We again use $M_i$ and $H$ to denote the resulting
expressions. They do not contain the variables $\tG^i$. Their
principal part is
\begin{eqnarray}
\label{BSSNHdef}
H&\simeq
&e^{-4\phi}\tg^{ij}\left(\tg^{kl}\tg_{ki,jl}-8\phi_{,ij}\right), \\
\label{BSSNMidef}
M_i&\simeq& \tA_{ij,k}\tg^{jk}-{2\over 3}K_{,i} .
\end{eqnarray}
The principal part of the evolution system for these constraints is
\begin{eqnarray}
\partial_0 H &\simeq& -2e^{-4\phi}\tg^{ij}M_{i,j}, \\
\partial_0 M_i &\simeq& \frac{1}{6} H_{,i}
+e^{-4\phi}\Bigl(
{a\over 2}\tg^{jk}G_{i,jk} \nonumber \\
&&+{a\over 6}\tg^{jk}G_{j,ik}
+\frac{1}{6}\tg^{jk}D_{,ijk}\Bigr), \\
\partial_0 G_i &\simeq & 2b M_i, \\
\label{Tdot}
\partial_0 {T}&\simeq& 0, \\
\label{Ddot}
\partial_0 D&\simeq &-2{T}.
\end{eqnarray}

If $T$ and $D$ vanish initially, they vanish at all times,
independently of the other three constraints. (This is true for the
full evolution equations, not shown here, as well as the principal
terms given in (\ref{Tdot},\ref{Ddot}).) From now on we restrict
attention to the subspace of solutions where these algebraic
constraints vanish. A discrete version of this restriction is
implemented in the ``BSSN'' code of most groups by projecting to
solutions of $T=D=0$ at each time step. It is not clear how exactly
this relates to the continuum restriction \cite{HinderPhD}. We shall
call the restricted continuum system BSSN-C (C is for
constrained). { Note that although NOR has 15 variables and
BSSN-C 17, two of those are algebraically redundant because we
consider $\tA_{ij}$ to be tracefree and $\tg_{ij}$ to have unit
determinant.}

There is a one-to-one correspondence between the {\em variables} of
BSSN-C and NOR given by (\ref{tgdef},\ref{phidef},\ref{tAdef}) and
\begin{equation}
\label{fGamma}
f_i=\tg_{ij}\tG^j+(4-6\rho)\phi_{,i}
\end{equation}
From (\ref{tgdef},\ref{phidef}) we also have the useful relations
\begin{equation}
d_i=\tg^{jk} \tg_{ij,k}+4\phi_{,i}, \qquad
t_i=12\phi_{,i},
\end{equation}
The {\em constraints} $H$, $M_i$ and $G_i$ { as defined for NOR in
(\ref{Hdef},\ref{Midef},\ref{Gdef}) and for BSSN in
(\ref{BSSNHdef},\ref{BSSNMidef},\ref{BSSNGdef}) are also equivalent in
the principal part modulo $T=D=0$}, and we have anticipated this by
using the same notation for both. Finally, if $c=-1/3$, $d=-a/3$ and
$\rho=2/3$, the principal parts of the {\em evolution equations} and
constraint evolution equations are also equivalent (for any $a$ and
$b$). The values of $a$ and $b$ in NOR then correspond to those of
BSSN-C, and we have anticipated this by using the same notation for
both.

{ In the following we focus on two versions of NOR: one where
$a=b=1$, $c=d=0$, which we shall call NOR-A, and one where $a=b=1$,
$c=d=-1/3$, which we shall call NOR-B. ADM is also considered in the
form of NOR with $a=b=c=d=0$. In these three cases we also assume
$\sigma=0$ unless specified otherwise.  Our results can be summarised
in the following}

{\bf Lemma 1:} {\em For any choice of lapse and shift, there is a
one-to-one algebraic correspondence between the variables of BSSN-C
and of NOR-B with $\rho=2/3$ and, using this, between the principal
parts of their evolution equations, constraints, and constraint
evolutions. In particular, BSSN-C is strongly/symmetric hyperbolic (in
the definition of \cite{reduct}) if and only if NOR-B is.}


\subsection{KST with fixed gauge}


BSSN-C and NOR-B are both second-order formulations and have the
same number of variables, namely 15. By contrast, the KST formulation
\cite{KST} is first-order and is not evidently the reduction to first
order of any second-order system \cite{bssn2}, while a first-order
reduction of NOR has 33 variables, not 30. Nevertheless, in this
Subsection we show that the two formulations are equivalent in the
precise sense that when only the principal part is considered, the KST
formulation of the Einstein equations with densitised lapse and fixed
shift \cite{KST} is an autonomous subsystem, in 30 of the 33
variables, of a reversible reduction to first order of NOR with
densitised lapse and fixed shift.  In the presence of an evolved lapse
and shift, this equivalence must be revisited because the evolution
equations for the lapse and shift could be written in a first-order
form which is not a reversible reduction of their second-order
form. This will be done in the following two Subsections.

In this Subsection, we shall assume that the lapse densitisation weight
$\sigma$ is a given parameter, and that the parameter $\rho$ of NOR is
also given. We shall prove that the remaining parameters $(a,b,c,d)$ of
NOR correspond to the parameters $(\zeta,\gamma,\chi,\eta)$
of KST as follows:
\begin{eqnarray}
\label{bcond}
4b&=&2\chi+4\eta-\rho(3\chi+\eta), \\
\label{cdef}
2c&=&\gamma, \\
\label{abdef}
4ab&=&-(1+3\sigma)\chi+\left({1\over2}-\sigma-{3\over2}\zeta\right)\eta,
\\
\label{bddef}
4db&=&-(1+2\gamma)\chi+\left({1\over 2}+{1\over
  2}\zeta+\gamma\right)\eta.
\end{eqnarray}
Note that $b$, $\chi$ and $\eta$ only appear in the combinations
$\eta/b$ and $\chi/b$.

We define the shorthand
\begin{eqnarray}
&& {\cal K}_{ij}(X_{klm},\zeta,\gamma,\sigma) \nonumber \\
&\equiv& 
-{1+\sigma\over 2}X_{(i|k \ \ ,|j)}^{\ \ \
  \ k}
-{1\over 2}{X_{kij}}^{,k} \br
+{1+\zeta\over 2}{X_{(ij)k}}^{,k}
+{1-\zeta\over 2}{{X_k}^k}_{(i,j)} \br
-{\gamma\over 2}
\g_{ij}\left({X_{kl}}^{l,k}-{{X_k}^{k,l}}_{,l}\right),
\end{eqnarray}
for any $X_{klm}$ that is symmetric in its last two indices.  With the
parameter identifications (\ref{cdef}-\ref{bddef}), ${\cal K}_{ij}$
obeys the identity
\begin{eqnarray}
&&{\cal K}_{ij}(X_{klm}+{\chi\over 2b}\g_{lm}Y_k+{\eta\over 2b}\g_{k(l}Y_{m)},
\zeta,\gamma,\sigma) \nonumber \\
\label{bigidea}
 &\simeq &
{\cal K}_{ij}(X_{klm},\zeta,2c,\sigma) 
+aY_{(i,j)}+d\g_{ij} Y_k{}^{,k}
\end{eqnarray}
for any $X_{klm}$ symmetric in its last two indices and $Y_i$. Here
$\simeq$ indicates that we have neglected first derivatives of
$\g_{ij}$. We now go from NOR to KST in three steps.


\subsubsection{First-order reduction of NOR} 


Using our shorthand notation, we can write the NOR evolution equations
with fixed gauge in the form
\begin{eqnarray}
\label{gdotfixedgauge}
\p_0 \g_{ij} &\simeq& -2K_{ij}, \\
\label{NORfdot}
\p_0 f_i &\simeq& -2{K_{ik}}^{,k}+\rho K_{,i}+2bM_i,\\
\p_0 K_{ij}&\simeq& {\cal K}_{ij}(\partial_k\g_{lm},*,2c,\sigma)
+aG_{(i,j)}+d\g_{ij} G_k{}^{,k},
\end{eqnarray}
where the $*$ indicates that the formal parameter $\zeta$ of $\cal K$
cancels because $\g_{ij,kl}=\g_{ij,lk}$. $G_i$ was defined above in
(\ref{Gdef}).

We reduce the NOR system to first order by introducing the auxiliary
variables $e_{kij}\doteq \g_{ij,k}$. The new evolution equations are
\begin{eqnarray}
\label{NOR1Kijdot}
\p_0 K_{ij}&\simeq &{\cal K}_{ij}(e_{klm},\zeta,2c,\sigma)
+a\bar G_{(i,j)} + d \g_{ij} \bar G_k{}^{,k}, \\
\label{edot}
\p_0 e_{kij} &\simeq& -2K_{ij,k},
\end{eqnarray}
with (\ref{gdotfixedgauge}) and (\ref{NORfdot}) as before. $\zeta$ now
arises as a reduction parameter, and we have used the shorthand $\bar
G_i \equiv f_i -{e^k}_{ki}-(\rho/2){e_{ik}}^{k}$ for the new
definition constraint.  Because no constraints have been introduced in
(\ref{edot}), the reduction is reversible in the sense defined in
\cite{reduct}. This means that the reduction is strongly (symmetric)
hyperbolic if and only if the original second-order system is strongly
(symmetric) hyperbolic. Furthermore, if both are
symmetric hyperbolic, their energies are related by identifying
$e_{kij}$ and $\g_{ij,k}$. { (See \cite{reduct} for definitions of
strong and symmetric hyperbolicity of second-order systems, and the
relation with their first-order counterparts.)} 

We now decompose $e_{kij}$ (18 components) into $\tilde e_{kij}$ (15
independent components) and $e_i$ (3 components) as follows:
\begin{eqnarray}
\label{defetilde}
\tilde e_{kij}&\equiv& e_{kij}-{\chi \over 2b}\,\g_{ij} e_k
-{\eta\over 2b}\,\g_{k(i}e_{j)}, \\
e_i &\equiv &{e^k}_{ki}-{\rho\over 2}{e_{ik}}^{k},\\
\label{etildetrace}
0&=& {\tilde e^k}_{\ ki}-{\rho\over 2}{\tilde e}_{ik}{}^{k} .
\end{eqnarray}
Here (\ref{bcond}) is necessary for (\ref{etildetrace}) to hold. 
The two parts of $e_{kij}$ evolve as 
\begin{eqnarray}
\label{etildedot}
\p_0 \tilde e_{kij}&\simeq& \p_0 e_{kij}-{\chi \over 2b}\,\g_{ij} \p_0
e_k
-{\eta\over 2b}\,\g_{k(i}\p_0 e_{j)},\\
\label{eidot}
\p_0 e_{i} &\simeq&-2{K_{ik}}^{,k}+\rho K_{,i}.
\end{eqnarray}

From (\ref{bigidea}) with $X_{kij}\to e_{kij}$ and $Y_i\to -e_i$
we see that (\ref{NOR1Kijdot}) can be rewritten as
\begin{eqnarray}
\label{NOR1Kijdotbis}
\p_0 K_{ij}&\simeq &{\cal K}_{ij}(\tilde e_{klm},\zeta,2c,\sigma)
+af_{(i,j)}+d\g_{ij} f_k{}^{,k},
\end{eqnarray}
The evolution equations of first-order NOR in final form are
(\ref{gdotfixedgauge},\ref{NORfdot},\ref{etildedot},\ref{eidot},\ref{NOR1Kijdotbis}).
When only the principal part is considered, the subsystem
(\ref{NORfdot},\ref{etildedot},\ref{NOR1Kijdotbis}) in the variables
$(\tilde e_{kij},f_i,K_{ij})$ is autonomous, and the variables
$(e_i,\g_{ij})$ follow passively, and so could, always in the
principal part approximation, be found after the subsystem has been
solved. Any estimate on the subsystem gives rises to a similar
estimate for the full system. Therefore the full system is strongly
hyperbolic if and only if the subsystem is. (We discuss symmetric
hyperbolicity in Step 3).


\subsubsection{From first-order NOR to KST}


We now define
\begin{eqnarray}
\label{ddef}
d_{kij}&\equiv& \tilde e_{kij}+{\chi \over 2b}\,\g_{ij} f_k
+{\eta\over 2b}\,\g_{k(i}f_{j)},\\
\label{ddef2}
&\equiv & e_{kij} +{\chi \over 2b}\,\g_{ij} \bar G_k
+{\eta\over 2b}\,\g_{k(i} \bar G_{j)}.
\end{eqnarray}
Note $d_{kij}$ is equal to $e_{kij}$ up to a constraint.
From (\ref{etildedot}) and (\ref{NORfdot}), or equivalently from
(\ref{edot}) and (\ref{Gdot}), its evolution equation is
\begin{eqnarray}
\label{KSTdkijdot}
\p_0 d_{kij} &\simeq& -2K_{ij,k} +\chi\g_{ij}M_k+\eta
\g_{k(i}M_{j)}.
\end{eqnarray}
Finally we use (\ref{bigidea}) with $X_{kij}\to \tilde e_{kij}$ and $Y_i\to
f_i$ to rewrite (\ref{NOR1Kijdotbis}) as 
\begin{eqnarray}
\label{KSTKijdot2}
\p_0 K_{ij} &\simeq& {\cal K}_{ij}(d_{klm},\zeta,\gamma,\sigma).
\end{eqnarray}
(\ref{gdotfixedgauge}), (\ref{KSTdkijdot}) and (\ref{KSTKijdot2})
together form the KST evolution equations for fixed gauge \cite{KST}.
(Note that $\sigma\equiv 2\sigma_{\rm KST}$ and that in
\cite{LindblomScheel} the notation $D_{kij}\equiv d_{kij}/2$ is used.)
In this second step we have discarded the dynamical variables
$e_i$ from the autonomous evolution system. Their place in $d_{kij}$
is taken by the $f_i$.

An equivalence between BSSN and a particular case of KST was
established in \cite{SarbachBSSN} without however clarifying the role
of the algebraic constraints in BSSN, or noting the special role of
the $e_i$.


\subsubsection{Symmetric hyperbolicity}


The conserved energy of NOR with fixed gauge in its original
second-order form is
\begin{eqnarray} \label{fixedNORenergy}
\epsilon(K_{ij},\g_{ij,k},f_i)&=&c_0\epsilon_0+c_1\epsilon_1
+c_2(d_i^2-\g_{ij,k}\g^{ik,j}) \nonumber \\
&& + c_3[G_i+b(d_i-t_i)]^2 ,
\end{eqnarray}
where $c_0\epsilon_0+c_1\epsilon_1$
is a complicated expression depending on
$(\rho,a,b,c,d,\sigma)$. (The special case $c=d=0$ is given in
\cite{bssn2}). The energy of NOR in first-order form is obtained by
replacing $\g_{ij,k}$ with $e_{kij}$, and setting
$c_2=(1+\zeta)/4$. In the first-order system $\zeta$ is a reduction
parameter which then determines $c_2$, while in the second-order
system $c_2$ arises as a free parameter in the energy.

The main difference between KST and first-order NOR is that the latter
has the additional variable $e_i$. However, there is a unique choice
for $c_3$ (a complicated expression involving $(a,b,c,d,\sigma)$ but
not $\rho$) which, together with $c_2=(1+\zeta)/4$, eliminates $e_i$
completely from the NOR energy. (This is non-trivial as three terms
$e^i e_i$, $e^i f_i$ and $e^i \tilde e_{ik}{}^k$ must be eliminated
from the energy with only the one parameter $c_3$ to adjust.) The
resulting NOR energy becomes the KST energy (compare App.~A of
\cite{LindblomScheel}) when $\tilde e_{kij}$ and $f_i$ are replaced by
$d_{kij}$ and its partial traces according to (\ref{ddef}) and with
the parameter identifications and choice of $c_3$ given
here. Then KST with fixed gauge is symmetric hyperbolic if and only if
NOR is. { We summarize our results in this Subsection in}

{\bf Lemma 2:} {\em There is a reduction to first order (from now,
NOR1) of the second-order NOR evolution equations (from now, NOR2)
with densitised lapse and fixed shift. A subsystem of NOR1 (from now,
NOR1a) is autonomous in the principal part approximation. It is
possible to relate the parameters of NOR1 and KST such that in the
principal part approximation there is a bijection between the
variables and evolution equations of NOR1a and KST. The evolution of the
variables $e_i$ which are in NOR1 but not in NOR1a is trivial in the
principal part approximation. This means that NOR1 is strongly
hyperbolic if and only if KST is. This in turn means that NOR2 is
strongly hyperbolic if and only if KST is.

Furthermore, NOR2 admits an energy which, when translated into
first-order form, does not contain $e_i$ and so is an energy for
NOR1a. There is bijection between this and the KST energy. This means
that NOR2 is symmetric hyperbolic if and only if KST is.}

{ (Here the hyperbolicity of second-order systems is the one defined in
\cite{reduct}.)}


\subsection{KST with evolved lapse: Sarbach-Tiglio}


The Sarbach-Tiglio (ST) formulation of the Einstein equations
\cite{SarbachTiglio} is KST with the evolved lapse
\begin{equation}
\label{Qdot}
\p_0(\ln \alpha)\simeq -\mu_L K, 
\end{equation}
where $\mu_L=2\sigma_{\rm eff}=\alpha^{-1}\p F/\p K$ in the notation
of \cite{SarbachTiglio}. We now show that the ST formulation is
(an autonomous subsystem of) a reversible reduction to first order of
NOR with this lapse. We start again from NOR and work towards ST.

The live lapse NOR evolution equations are (\ref{gdotfixedgauge}) and
(\ref{NORfdot}) together with
\begin{eqnarray}
\label{NORlivelapseKijdot}
\p_0 K_{ij}&\simeq& -(\ln \alpha)_{,ij}+{\cal K}(g_{ij,k},*,2c,0) \br
+aG_{(i,j)}+d\g_{ij} G_k{}^{,k}.
\end{eqnarray}
We introduce the auxiliary variable
\begin{equation}
a_i \doteq (\ln \alpha)_{,i}
\end{equation}
and evolve it as
\begin{equation}
\p_0 a_i\simeq -\mu_L K_{,i}, 
\end{equation}
so that its introduction is reversible. In analogy with (\ref{ddef2})
we now define
\begin{equation}
\label{Aitrick}
A_i\equiv a_i+{\xi\over 2b} G_i
\end{equation}
which obeys 
\begin{equation}
\label{Aidot}
\p_0 A_i\simeq -\mu_L K_{,i} +\xi M_i.
\end{equation}
with $\xi$ a new parameter. We complete the reduction to first order
as in the fixed gauge case by introducing $e_{kij}$, then replacing
its components $e_i$ by $f_i$ to obtain $d_{kij}$. This transforms
(\ref{NORlivelapseKijdot}) into
\begin{eqnarray}
\label{STKijdot}
\p_0 K_{ij} &\simeq& -A_{(i,j)}+{\cal K}(d_{kij},\zeta,\gamma,0).
\end{eqnarray}
(\ref{gdotfixedgauge}), (\ref{KSTdkijdot}), (\ref{Aidot}) and
(\ref{STKijdot}) form the ST evolution equations, where the NOR and
KST parameters are related by (\ref{bcond}), (\ref{cdef}),
(\ref{bddef}) with (\ref{abdef}) replaced by
\begin{eqnarray}
\label{abdefST}
4ab&=&-\chi+\left({1\over2}-{3\over2}\zeta\right)\eta-2\xi.
\end{eqnarray}
Compared to (\ref{abdef}), we have set $\sigma=0$, and the additional
term $-2\xi$ appears on the right-hand side because of the appearance
of $(2\xi/b) G_{(i,j)}$ in $A_{(i,j)}$. We have shown that ST is
equivalent to a first-order reduction of NOR with the lapse evolution
(\ref{Qdot}).  The characteristic speeds squared of \cite{SarbachTiglio}
simplify in our notation to $\lambda_1=\mu_L$,
$\lambda_2=1+4(c-bd)$ and $\lambda_3=ab$.

Now hold $\mu_L$ fixed and set $\sigma=0$. Then NOR has 5
parameters $(a,b,c,d,\rho)$ and KST has 5 parameters
$(\gamma,\zeta,\chi,\eta,\xi)$, but there are only 4 algebraic
relations between them, namely $2c=\gamma$, and (for generic
values of $\rho$) three linear relations between $(b,ab,bd)$ and
$(\chi,\eta,\xi)$. One can consider $\rho$ in NOR and $\zeta$ in KST as
arbitrary in their system with no counterpart in the other evolution
system.

ST find a 3-parameter family of strongly hyperbolic formulations whose
NOR equivalent is $ab=1$, $d=ac$. Their 2-parameter family is
the special case $c=-1/4$. The 3 free parameters can be taken to be
$(\gamma,\zeta,\eta)$ in KST and $(\rho,a,c)$ in NOR, with only 2
algebraic relations between these sets. ST also find a 2-parameter
family of symmetric hyperbolic formulations, which corresponds to NOR
with $c=-1/3$, $d=-7/30$. The free parameters can be taken to be
$(a,b,\rho)$ on the NOR side but on the KST side there are only two,
for example $(\eta,\xi)$, which are determined by $(a,b)$ alone.

Note finally that the equivalence between NOR and KST would hold
for any gauge that specifies the first time derivative of the lapse,
with fixed shift.

{\bf Lemma 3:} {\em Lemma 2 holds also for NOR and KST with a
Bona-Masso type evolved lapse and fixed shift.}


\subsection{KST with evolved lapse and shift: Lindblom-Scheel}


The formulation of Lindblom and Scheel (LS) \cite{LindblomScheel} is
again based on KST, but with an evolved shift as well as lapse. The
gauge conditions are (\ref{LS1}-\ref{LS2}) below, but the following
comments apply to any gauge where the first time derivatives of the
lapse and shift are specified. Besides the auxiliary variables
$d_{kij}$ and $A_i$ already present in ST (called $2D_{kij}$ and $T_i$
in LS), the auxiliary variable ${M_k}^i\doteq \alpha^{-1}
{\beta^i}_{,k}$ is introduced, and all possible constraints with the
right number of indices and level of derivatives are added to the
right-hand side of its evolution equation. In our notation, these are
\begin{equation}
\p_0 {M_k}^i = \dots +
\tilde\psi_4{G_k}^{,i}+\tilde\psi_5{G^i}_{,k}
+\tilde\psi_6{\delta_k}^i{G_l}^{,l}
+\tilde\psi_7{\delta_k}^i H,
\end{equation}
plus auxiliary constraints. (Our constants $\tilde
\psi_{4\dots 7}$ are linear combinations of $\psi_{4\dots 7}$ of
LS. The derivatives of $G_i$ are added not in this explicit form but
implicitly with the definition constraints of $d_{kij}$ and $A_i$. We
think of $G_i$ as the relevant part of those constraints, and the
definition constraints of $a_i$ and $e_{kij}$ as the part that is
irrelevant because its addition does not change the reversibility of
the first-order reduction.)  Because there are no constraints whose
time-derivatives are equal to these constraint terms, we cannot use a
trick analogous to (\ref{Aitrick},\ref{Aidot}) to recreate the effect
of these constraint additions in the first-order reduction within
fully 2nd-order NOR. The smallest system fully equivalent to LS that
{\it could} be constructed would be a hybrid between second-order NOR
and LS comprising the NOR variables $(\g_{ij},K_{ij},f_i)$ and gauge
variables $(\alpha,\beta^i)$ but also ${M^i}_k$. We have not analysed
this further.

Conversely, the introduction of ${M^i}_k$ remains reversible if
we set $\tilde\psi_{4\dots7}=0$. In the notation of LS, this
corresponds to the conditions 
\begin{eqnarray}
\label{conds1}
16 \chi\xi \psi_3 + (8\chi^2+8\chi\eta+3\eta^2)\psi_4 &=& 0, \\
\eta\psi_4 - 4\chi \psi_5 &=& 0, \\
\psi_5 -\psi_6 &=& 0, \\
\label{conds4}
\psi_7 &=& 0
\end{eqnarray}
on $\psi_{4\dots7}$. With these restrictions, LS is a reversible
first-order reduction of NOR with the same gauge. The identification
between NOR and LS parameters is then given by (\ref{bcond}),
(\ref{cdef}), (\ref{bddef}) and (\ref{abdefST}) and $\psi_1=2\xi$. The
remaining free parameters $(\psi_2,\psi_3,\psi_8,\psi_9,\psi_{10})$ of
LS are reduction parameters connected to the introduction of auxiliary
variables.  They have no direct counterpart in the NOR evolution
equations but can be identified with free parameters in the NOR
energy. 

{\bf Lemma 4:} {\em Lemma 2 holds also for NOR with a
Lindblom-Scheel-type evolved lapse and shift and the subset of LS formulations
characterised by the parameter restrictions (\ref{conds1}-\ref{conds4}).}


\section{Analysis of live gauge conditions}
\label{section:gauges}\subsection{Symmetry-seeking coordinate choices}


In the 3+1 approach the gauge freedom of general relativity is fixed
by specifying the lapse $\alpha$ and shift $\beta$ in terms of the
3-metric $\g_{ij}$ and the extrinsic curvature $K_{ij}$, and
possibly given functions of $(t,x^i)$, through algebraic or
differential equations. 

A unifying viewpoint for the classification of gauge conditions was given
in \cite{symmcoord}. A good gauge choice should have the property that
if the spacetime is stationary the vector field $(\p/\p t)^a=\alpha
n^a+\beta^a$ becomes equal to the Killing vector, {\em whatever the
initial slice and the spatial coordinates on it are}. We do not
require that $(\p/\p t)^a$ be timelike but only that $\alpha = -n_a
(\p/\p t)^a>0$, where $n_a$ is the unit normal on the surfaces of
constant $t$. This is assumed in the following. In such
coordinates, $\g_{ij}$ and $K_{ij}$, as well as $\alpha$ and $\beta^i$
become independent of $t$. Such conditions can be formulated in terms
of the vanishing of 3+1 time derivatives. Modifying the
terminology of \cite{symmcoord} we shall use the term ``freezing
conditions'' for gauge conditions in which the metric is {\it
immediately} time-independent in the presence of a Killing vector, and
restrict the term ``symmetry-seeking'' for gauge conditions in which
the metric becomes time-independent {\it asymptotically}. 


\subsection{Elliptic freezing conditions}


If the dynamical variables are frozen in the presence of a Killing
vector, so are the lapse and shift. Freezing conditions for the lapse
and shift can therefore not contain their time derivatives. An example
is the $K$-freezing lapse
\begin{equation}
\dot K=0.
\end{equation}
For given $\beta^i$, $\g_{ij}$ and $K_{ij}$, this is an elliptic
equation for $\alpha$, whose principal part is the Laplace operator
associated with $\g_{ij}$. It is clear from the way we have written it
that this is a freezing coordinate condition. Irrespective of the
evolution system, we always use the simplest form
\begin{equation}
\label{Kdotgauge}
\dot K = \beta^i K_{,i} - \Delta \alpha + \alpha K_{ij}K^{ij} 
\end{equation}
in any gauge condition. The principal part of this can be written as
\begin{equation}
\label{Kdotgaugepp}
\p_0 K \simeq - \p_i\p^i (\ln\alpha).
\end{equation}

If $K=0$ in the original slice (and therefore on all slices), the
$K$-freezing lapse is called maximal slicing. Maximal slicing
combined with minimal strain shift
\begin{equation}
D^i \dot \g_{ij}=0,
\end{equation}
or minimal distortion shift
\begin{equation}
D^i \dot {\tg}_{ij}=0,
\end{equation}
were first suggested for use in numerical relativity in \cite{SmarrYork}.

If, neglecting lower-order terms, we pull the time derivative out of
the divergence, we can integrate these last two gauge conditions over
$t$. In the context of the BSSN formulation, the gauge condition
$\dot{\tG^i}=0$ has been considered under the name $\Gamma$-freezing
shift. We consider a generalisation in the form of the 2-parameter
family of shift conditions
\begin{equation}
\dot {\bar f^i}=0, 
\end{equation}
where
\begin{eqnarray}
\bar f_i &\equiv& d_i-{\bar\rho\over 2}t_i+sG_i \\
\label{fbardef}
&=&sf_i+(1-s)d_i+{s\rho-\bar\rho\over 2}t_i, \\
&=&s \tg_{ij}\tG^j+(1-s)d_i+\left(\frac{s}{3}-\frac{\bar\rho}{2}\right) t_i,
\end{eqnarray}
where $\bar\rho$ and $s$ are constant parameters. Essentially, $\bar
f^i$ is the divergence of the 3-metric. The price we pay for pulling
the time derivative outside the divergence is that while $\dot\g_{ij}$
transforms as a tensor under spatial coordinate transformations, $\bar
f^i$ does not. 
(However, $\bar f^i=0$ and similar conditions
(Dirac gauge) can be formally made covariant by writing $f^i$ as
the covariant divergence of $\g_{ij}$ with respect to a flat background
metric \cite{Meudon}).
$\bar\rho$ parameterises a conformal weight, and $s=1$ gives us the
option to express the gauge condition in terms of the NOR variable
$f^i$, while with $s=0$ the gauge condition is expressed in terms of
the 3-metric. With $s=1$, $\rho=\bar\rho=2/3$ we have $\bar
f^i=\g^{-1/3}\tG^i$.

We shall loosely call all these gauge conditions
$\Gamma$-freezing.  They are elliptic in $\beta^i$ for $\bar\rho<2$,
and by construction are freezing. The principal part of $\dot {\bar
f^i}=0$ can be read off from
\begin{eqnarray}
\label{fbardotpp}
\p_0 {\bar f_i} &\simeq& \alpha^{-1}\g_{ik}{{\beta^k}_{,j}}^{,j}+
(1-\bar\rho) \alpha^{-1}{\beta^j}_{,ji} \br
+2(sb-1){K_{ij}}^{,j}+(\bar\rho-2sb) K_{,i}.
\end{eqnarray}
The principal part in $\beta^i$ coincides with minimal strain shift
and minimal distortion shift  for $\bar\rho=0$ and $\bar\rho=2/3$,
respectively. 


\subsection{Parabolic drivers}


Coordinate drivers turn elliptic equations for the lapse or shift into
heat equations or wave equations by adding one or two time
derivatives. Their motivation is that in a spacetime with a Killing
vector, and with suitable boundary conditions, we expect that the
solution of the resulting heat or wave equation will tend towards a
(time-independent) solution of the original elliptic freezing
condition.

Parabolic drivers for the lapse and shift were, to our knowledge,
first suggested in \cite{Balakrishna}. The basic (explicit) parabolic
``$K$-driver'' is
\begin{equation}
\label{dotalphadotK}
(\ln\alpha)\dot{}=-\nu_L\dot K
\end{equation}
for some constant $\nu_L>0$.  
The choice of powers of $\alpha$ here
and in the following is determined by the requirement that the gauge
choice be invariant under the rescaling of $t$ by a constant
factor. The equation is also invariant under arbitrary changes of the
spatial coordinates.

The elliptic equation $\dot K=0$ has been turned into a heat equation
for $\alpha$. If all variables other than $\alpha$ were
time-independent, the solution of the parabolic equation would tend to
a solution of the elliptic equation $\dot K=0$, but in reality all
variables evolve together.  Whether the solution actually tends
towards a freezing condition will therefore depend on the physical
problem. However, we can see that the driver is at least compatible
with being symmetry-seeking in the sense that the evolution of
$\alpha$ stops if and only if the freezing condition $\dot K=0$ is
obeyed.

Integrating explicitly over $t$, we obtain the basic implicit
parabolic $K$-driver
\begin{equation}
\label{implicitparaboliclapse}
\ln\alpha=-\nu_L K+\phi(x).
\end{equation}
As $K$ is a dynamical variable, this is now an algebraic gauge
condition rather than a heat equation, and so the implicit and
explicit forms are not equivalent as PDE systems. However, if the
constraints are obeyed the two forms will generate the 
same gauge. In particular, $\phi(x)$ allows us to still set gauge
initial data for $\alpha$ independently of the value of $K$ given by
the geometric initial data, as
\begin{equation}
\label{initiallapse}
\phi(x)=(\nu_L K+\ln\alpha)_0.
\end{equation}
(In the remainder of this Section the suffix 0 denotes the initial
value of a quantity.) On the other hand, if the parabolic driver
succeeds in driving $K$ to a time-independent value, $\phi(x)$ also
characterises a condition on the asymptotic values of $K$ and
$\alpha$, namely
\begin{equation}
\label{finallapse}
\phi(x)=(\nu_L K+\ln\alpha)_\infty,
\end{equation}
where the suffix $\infty$ denotes the asymptotic value. Such a
condition is to be expected: while elliptic conditions freeze $K$ and
$\bar f^i$ at whatever value they have in the initial data, a driver
condition freezes $K$ and $\bar f^i$ only asymptotically and at values
which depend not only on the geometric initial data $K_{ij}$ and
$\g_{ij}$, but also on the initial values of $\alpha$ and $\beta^i$
(and $\dot\alpha$ and $\dot\beta^i$ in the case of a hyperbolic
driver, see below).

The situation for the $\Gamma$-freezing shift conditions is very
similar to the $K$-freezing lapse condition. The basic parabolic
``$\Gamma$-driver'' is
\begin{equation}
\label{parabolicGammadriver}
\dot \beta^i = \nu_S (\alpha\bar f^i)\dot{}
\end{equation}
for some constant $\nu_S$.
The left-hand side is a vector and the right-hand side is not
(which means that the time evolution does not commute with generic
spatial coordinate transformations), but the
equation transforms correctly under rescaling the spatial coordinates
by a constant factor and under rotations. The factor of $\alpha$
makes the equation invariant under rescaling $t$ by a constant
factor. For $\bar\rho=0$ this is a heat equation for $\beta^i$, and
for $\bar\rho<2$ it is parabolic. Again, this condition is compatible
with being symmetry-seeking. Similarly, an implicit parabolic shift
condition is
\begin{eqnarray}
\label{implicitparabolicshift}
\beta^i&=&\nu_S\alpha \bar f^i-\phi^i(x).
\end{eqnarray}

What would happen if we replaced $\p_t$ with $\p_0$? The
straightforward replacement,
\begin{equation} 
\label{d0straight}
\p_0 (\ln\alpha)=-\nu_L \p_0 K,
\end{equation}
can be integrated using the method of lines:
\begin{equation}
\ln\alpha+\nu_L K=\phi[\tilde x(x,t)],
\end{equation}
where 
\begin{equation}
\dot{\tilde x^i}=\beta^j{\tilde x^i}_{,j}, \qquad \tilde x^i(x,0)=x^i,
\end{equation}
and $\phi(x)$ is given by (\ref{initiallapse}). Clearly this couples
the lapse and shift conditions through $\tilde x^i$, leading to a
gauge condition which is different from the original one. However, if
we write
\begin{equation} 
\label{d0plussource}
\p_0 (\ln\alpha)=-\nu_L \p_0 K+\alpha^{-1}\beta^i\p_i\phi(x),
\end{equation}
this can be rewritten as the $\p_0$ derivative of
(\ref{implicitparaboliclapse}). If we again define $\phi(x)$ from
(\ref{initiallapse}), we have (\ref{implicitparaboliclapse})
identically. Taking a $\p_t$ derivative, we obtain
(\ref{dotalphadotK}). Therefore (\ref{dotalphadotK}) and
(\ref{d0plussource}) give rise to the same gauge (when all Einstein
equations are obeyed). In particular we have
(\ref{finallapse}). However, the $\p_0$ and $\p_t$ implementations
have different principal parts, and so their well-posedness is not
obviously equivalent. The use of $\p_0$ in the gauge conditions
simplifies the well-posedness analysis, but $\p_0 \beta^i$ is similar
to the nonlinear term in Burger's equation, and may develop
shocks \cite{ReimannAlcubierre}. Contrary to the claim in
\cite{LindblomScheel}, this problem cannot be fixed simply by
introducing an auxiliary variable for the first derivatives of the
shift, as the reduction admits the same blow-up solutions.


\subsection{Hyperbolic drivers}


Hyperbolic drivers were discussed in \cite{AEIcoordinates}. The basic
explicit hyperbolic $K$-driver is
\begin{equation}
\left(\alpha^{-1}(\ln\alpha)\dot{}\,\right)\dot{}+2\kappa_L(\ln\alpha)\dot{}
=-\mu_L\dot K, 
\end{equation}
for some $\mu_L>0$ and $\kappa_L>0$. This equation has the same
transformation properties as the parabolic lapse driver. The
motivation is that the elliptic equation $\dot K=0$ will be solved by
turning it into a wave equation with a friction term for
$\alpha$. Once again, this would be true if all variables other than
$\alpha$ were time-independent but it is not clear if this
gauge condition is symmetry-seeking in general when it is applied to a
situation where all dynamical variables evolve.

As we have chosen the outermost derivative to be $\p_t$, we can again
integrate in $t$ to obtain the basic implicit hyperbolic $K$-driver
\begin{eqnarray}
\dot\alpha=-\alpha^2 \mu_L K-2\kappa_L\alpha^2\ln\alpha+\alpha^2 \phi(x),
\end{eqnarray}
which is again equivalent as a gauge condition to the explicit driver
as long as the constraints are obeyed. $\phi(x)$ can be used to set
initial data for $\dot\alpha$ independently of $\alpha$ and $K$, but
if the spacetime admits a Killing vector and the driver is indeed
symmetry-seeking then it is also related to the asymptotic values of
$\alpha$ and $K$:
\begin{equation}
\label{finalimplicithyperboliclapse}
\phi(x)=(\mu_L K+2\kappa_L\ln\alpha)_\infty.
\end{equation}

Another damping mechanism (suggested by Shibata \cite{Shibata} for the shift
driver) for an implicit hyperbolic driver is
\begin{eqnarray}
\label{Shibatalapse}
\dot\alpha&=&-\mu_L\alpha^2K-2\nu_L\alpha\dot K+\alpha^2 \phi(x).
\end{eqnarray}
By linearising in $\alpha$, holding all other variables fixed and
carrying out a Fourier analysis, one sees that all modes are
damped. However, this linear equation for $\alpha$ (with all other
variables held fixed) is neither hyperbolic nor parabolic. Shibata
chooses $\eta\propto \Delta x$, so that in the continuum limit the
implicit hyperbolic driver is recovered. One advantage of this damping
mechanism is that now $\phi(x)=\mu_L K_\infty$, giving better control
of the final gauge.

Again, a similar analysis applies to the $\Gamma$-freezing shift
conditions. The basic explicit hyperbolic $\Gamma$-driver (with
damping) is
\begin{equation}
(\alpha^{-1}\dot\beta^i)\dot{} +2\kappa_S\dot \beta^i= (\mu_S\alpha
\bar f^i)\dot{}.
\end{equation}
(This equation has the same transformation
properties as the parabolic shift driver.)
For $\bar\rho=0$ this is a wave equation for $\beta^i$. 
The equivalent implicit hyperbolic $\Gamma$-driver is
\begin{eqnarray}
\dot \beta^i&=&\mu_S\alpha^2\bar f^i-2\kappa_S\alpha\beta^i-\alpha
\phi^i(x), 
\end{eqnarray}
and the alternative damping mechanism is
\begin{eqnarray}
\label{Shibatashift}
\dot \beta^i &=& \mu_S\alpha^2 \bar f^i + 2\nu_S \alpha \dot{\bar f^i} -\alpha
\phi^i(x),
\end{eqnarray}
the latter obeying $\phi^i(x) = \mu_S\alpha \bar f^i_\infty$.

Finally we note that the explicit parabolic or hyperbolic drivers
differ from the explicit ones in that in the expressions for $\dot K$
or $\dot {\bar f^i}$ that appear on their right-hand side one can add
suitable constraints in order to change the principal part of the
equations, while this is not possible in the implicit case. 

We can strip the outer time derivative off the explicit hyperbolic
drivers to obtain the implicit hyperbolic drivers even if that outer
time derivative is $\p_0$ with the same trick of adding a source term
we used above in (\ref{d0plussource}). This affects the nature and
well-posedness of the system, but leaves the gauge choice unaffected.
This is not true if we replace the remaining (inner) time derivative
$\p_t$ with $\p_0$ on the left-hand side of the implicit hyperbolic
drivers. For example, with
\begin{equation}
\p_0(\ln\alpha)=-\mu_L K-2\kappa_L\ln\alpha+\phi(x),
\end{equation}
(\ref{finalimplicithyperboliclapse}) is replaced by
\begin{eqnarray}
\phi&=&(\mu_LK+2\kappa_L\ln\alpha+\alpha^{-2}\beta^i\alpha_{,i})_\infty,
\end{eqnarray}
which couples the asymptotic lapse and shift.


\subsection{Boundary conditions}


Elliptic gauge conditions guarantee that 4 dynamical variables are
frozen in any spacetime, but $\g_{ij}$, $K_{ij}$, $\alpha$ and
$\beta^i$ are all frozen in a stationary spacetime only if the
boundary conditions are compatible with the Killing vector.
Finding boundary conditions which are at the same time compatible with the
symmetry-seeking gauge, the constraints, and give rise to a well-posed
initial-boundary value problem is an
important problem that has not yet been studied in any depth. The
parabolic and hyperbolic drivers require the same type of boundary
conditions as the elliptic freezing conditions, namely Dirichlet,
Neumann or mixed boundary conditions for the lapse and shift. These
cannot be completely homogeneous, as they must fix an overall common
factor in the lapse and shift (corresponding to the freedom to rescale
$t$). We shall not otherwise investigate this issue here.

Finally, we note that when the the gauge evolution uses $\p_0$ and a
symmetric hyperbolic evolution system is obtained, one can arrange the
characteristic speeds so that singularity excision surfaces inside
black holes become purely outflow, and no boundary conditions are
required. When the hyperbolic gauge evolution uses $\p_t$, or when the
gauge condition is elliptic or parabolic, boundary conditions for the
gauge are required at excision boundaries. This may be 
desirable or undesirable. From the fact that the Einstein equations
are well-posed in harmonic gauge, where excision surfaces are purely
outflow, it is clear that no physical information can leave the black
hole even if boundary conditions are imposed inside.


\section{Hyperbolicity with implicit hyperbolic lapse and shift drivers}
\label{section:hyperbolicity}


\subsection{Generalised Lindblom-Scheel gauge}

In the following we investigate the strong and symmetric hyperbolicity
of ADM, NOR-A and NOR-B coupled to a family of implicit hyperbolic
driver gauge conditions. These conditions are
\begin{eqnarray}
\label{LS1}
(\p_0+\zeta_L\p_{\bar\beta})(\ln\alpha)&\simeq & 
-\mu_L K +\epsilon_L \alpha^{-1} {\beta^i}_{,i}, \\
\label{LS2}
(\p_0+\zeta_S\p_{\bar\beta})\beta^i&\simeq &
\mu_S \alpha \bar f^i+\epsilon_S \alpha (\ln\alpha)^{,i}, 
\end{eqnarray}
where we have not written the friction damping terms and the
integration constants $\phi,\phi^i$ because they are non-principal. We
have introduced the shorthands $\bar \beta^i\equiv \alpha^{-1}\beta^i$
and $\p_{\bar\beta}\equiv \bar\beta^i\p_i$.

The free parameters in the principal part of the gauge conditions are
$\bar\rho$, $\zeta_L$, $\zeta_S$, $\mu_L$, $\mu_S$, $\epsilon_L$ and
$\epsilon_S$. With $\zeta_L=\zeta_S=0$ these are the conditions
considered by Lindblom and Scheel \cite{LindblomScheel} and the
parameters $\mu_L$, $\mu_S$, $\epsilon_L$ and $\epsilon_S$ have the
same definition. The parameter $\bar\rho$ is related to the parameter
$\lambda$ of \cite{LindblomScheel} by
$\bar\rho=1+\lambda$. $\zeta_L=\zeta_S=0$ simplifies the analysis of
hyperbolicity, while $\zeta_L=\zeta_S=1$ corresponds to the hyperbolic
drivers we have discussed above and gives rise to simpler asymptotic
gauge conditions.  Harmonic gauge is the special case
$\zeta_L=\zeta_S=0$, $\mu_L=\mu_S=1$, $\epsilon_L=0$, $\epsilon_S=-1$,
$\bar\rho=1$. We shall refer to (\ref{LS1},\ref{LS2}) in generality as
LS gauge.

The parameter $s$ appears implicitly in the gauge conditions, but its
value does not affect the gauge as long as the constraints are obeyed,
and it should therefore be considered as a parameter of the
formulation. Both NOR-A and NOR-B have the free parameter $\rho$, but
in the following examples neither the characteristic speeds nor the
well-posedness depend on the value of $\rho$.


\subsection{Using $\p_0$}


We now follow \cite{LindblomScheel} and investigate
(\ref{LS1},\ref{LS2}) with $\zeta_L=\zeta_S=0$, thus using $\p_0$ as the time
derivative.  In the principal part, frozen coefficient approximation
the equations decouple into three blocks, which we consider in
turn. The statements below hold for ADM, NOR-A and NOR-B, unless
further qualified. 

The transverse traceless symmetric tensor block of the principal part
is always diagonalisable, with $\lambda=\pm 1$ (twice). The transverse
vector block has eigenvalues $\lambda=\pm 1$ (for NOR) or 0 (twice,
for ADM), plus $\lambda=\pm\sqrt{\mu_S}$. It is diagonalisable for
$\mu_S>0$ in the ADM case, and for $\mu_S>0$ with either $\mu_S\ne 1$
or $s=1$ in the NOR-A and NOR-B cases.

The scalar block always has $\lambda=0,\pm 1,\pm v_{\pm}$, where 
\begin{eqnarray}
v_{\pm}^2&\equiv& A \pm B, \label{sol1} \\
2A &\equiv & \mu_L + (2-\bar\rho)\mu_S+\epsilon_S\epsilon_L , 
\label{sol2} \\
B^2 - A^2 &\equiv & (2-\bar\rho)(\epsilon_L-\mu_L)\mu_S. \label{sol3}
\end{eqnarray}
$v_\pm$ depend only on the gauge choice, not the parameters $\rho$,
$s$ of the formulation.  Defining $B$ to be positive, they are real if
and only if $0\le B \le A$. At generic points in the interior of this
region the scalar sector is diagonalisable because the four gauge
speeds are different. On the lines $A=B$, $B=0$ and $A-B=1$, $A+B=1$
some of the characteristic speeds are repeated, and generically these
marginal cases are not diagonalisable. In particular, $A=B$
(i.e. $v_-=0$) happens for $\bar\rho=2$ or $\mu_S=0$ or
$\epsilon_L=\mu_L$ and then the system is never diagonalisable. There
are further parameter choices where some of the marginal cases become
diagonalisable, but we do not analyse them here.  We just mention that
under the asumption $A\ne B$ and using the fact that for ADM, NOR-A
and NOR-B we have $c-bd=0$, diagonalisability of the whole $7\times 7$
scalar sector with eigenvalues $(0,\pm 1,\pm v_-, \pm v_+)$ is
equivalent to diagonalisability of the matrix
\begin{equation} \label{redmatrix}
\left(\matrix{
\epsilon_L\epsilon_S+(2-\bar\rho)\mu_S & -\epsilon_L & 0 \cr
-\mu_L\epsilon_S-(2-\bar\rho)\mu_S & \mu_L & 0 \cr
-\mu_L\epsilon_S-(2bs-\bar\rho)\mu_S & 1-2ab+\mu_L & 1}\right)
\end{equation}
with eigenvalues $(1,v_-^2,v_+^2)$.

The results simplify if we partially decouple the lapse and shift
conditions by setting $\epsilon_L=0$. Then $v_\pm$ are simply
$\sqrt{\mu_L}$ and $\sqrt{(2-\bar\rho)\mu_S}$ and the matrix
(\ref{redmatrix}) becomes triangular, which allows a simple discussion
of its diagonalisability: as we said, we always need $\mu_L>0$,
$\mu_S>0$, $\bar\rho<2$ to avoid a zero eigenvalue $v_-^2$. If the
three eigenvalues are different then diagonalisability is guaranteed
for both ADM and NOR. If $\mu_L=1$ then we need $ab=1$, which is only
possible for NOR; if $(2-\bar\rho)\mu_S=\mu_L$ we need
$\epsilon_S=-1$; finally if $(2-\bar\rho)\mu_S=1$ we need
$(\mu_L\epsilon_S+1)(1-ab)=\mu_S(bs-1)(\mu_L-1)$, which requires
$bs=1$ for NOR-A/B. It is possible to make all scalar speeds
equal to $\pm 1$ using the intersection of those three cases:
$\epsilon_L=0$, $\epsilon_S=-1$, $ab=bs=\mu_L=1$, $c=bd$ and
$(2-\bar\rho)\mu_S=1$, while $s,d,\rho,\bar\rho,\sigma$ still are free
parameters. We can set all speeds (including vector speeds)
equal to zero or one by further setting $\mu_S=s=1$, which then
requires $\bar\rho=1$. The result is the implicit form of harmonic
gauge in NOR with $a=b=1$ and $c=d$, while $d,\rho,\sigma$ are still
free parameters.  For example, setting $d=-1/3$, $\rho=2/3$ and
$\sigma=0$ we obtain a system equivalent to BSSN-C with implicit
harmonic gauge. (Note that although the gauge is harmonic gauge, this
system is not equivalent to the usual harmonic evolution system.)

In all other cases we cannot make all speeds equal to zero or one.
This is not necessarily bad: one may want constraint-violating modes to
have speeds less than 1 so that they do not pile up on the black-hole
horizon \cite{LindblomScheel}, and one may want gauge modes to have
speeds larger than 1 so that one can impose boundary conditions on
them at the black hole excision boundary (inside the horizon) in order
to control the gauge better \cite{Lindblompersonal}. We have not
investigated these issues here. A simple example is
\begin{eqnarray}
\label{example}
\quad \bar\rho=2/3, \quad \mu_L=\mu_S={1\over 4}, 
\quad \epsilon_L=\epsilon_S=0,
\end{eqnarray}
which has speeds $\lambda=(0,\pm 1,\pm 1/2,\pm 1/\sqrt{3})\le 1$,
independently of $\rho$ and $s$. 

Finally we note that the lapse driver alone, with fixed shift, is
strongly hyperbolic with NOR-A/B for $\mu_L>0$, but not with ADM.


\subsection{Using $\p_t$}


We now set $\zeta_L=\zeta_S=1$, which is equivalent to replacing $\p_0$ by
$\alpha^{-1}\p_t$ as the time derivative on the left-hand side of
(\ref{LS1}-\ref{LS2}). As before all speeds $\lambda$ are with respect
to normal observers and in units of the speed of light. 

The characteristic speeds in the vector sector are $\lambda=\pm 1$
(for NOR) or 0 (twice, for ADM), plus
\begin{equation}
\lambda=-{\bar\beta_n\over 2}\pm \sqrt{\left({\bar\beta_n\over
    2}\right)^2+\mu_S}.
\end{equation}
With ADM, the vector sector is diagonalisable for $\mu_S>0$, for any
$\bar\beta_n$. With NOR, $\mu_S>0$ is also necessary, but not
sufficient: two speeds which are generically distinct coincide for
$\mu_S-1=\pm \bar\beta_n$, and we need to impose $\mu_S-1\ne\pm
\bar\beta_n$ to maintain diagonalisability. 

The fact that this condition depends on
$\bar\beta_n=\bar\beta^i\omega_i/|\omega|$
raises a serious problem: As long as $|\bar\beta|\equiv
\sqrt{\bar\beta^i\bar\beta_i}>|\mu_S-1|$, this inequality will be
violated for all $\omega_i$ that lie on a cone in wave number
space. In particular, it is then violated for arbitrarily large
$|\omega|$. This means that for sufficiently large shift, the system
is only weakly hyperbolic. The inequality required for strong
hyperbolicity is therefore
\begin{equation}
\label{shift0}
|\bar\beta|<|\mu_S-1|.
\end{equation}
Numerical evolutions should be unstable when it is violated.
However for $s=1$ (\ref{shift0}) is not required.

The characteristic speeds in the scalar sector are $\lambda=0,
\pm 1$ and the four roots of the quartic equation
\begin{eqnarray}
 \lambda^4
&+& (\zeta_L+\zeta_S)\bar\beta_n \lambda^3 \nonumber \\
&+& \left[\zeta_L\zeta_S\bar\beta_n^2-\epsilon_L\epsilon_S+
  (\bar\rho-2)\mu_S-\mu_L\right] \lambda^2 \nonumber \\
&+& \left[(\bar\rho-2)\zeta_L\mu_S-\zeta_S\mu_L\right]\bar\beta_n\lambda
\nonumber \\
&+& (\epsilon_L-\mu_L)\mu_S(\bar\rho-2)
= 0, 
\end{eqnarray}
For $\zeta_L=\zeta_S=0$ this is a biquadratic equation and its
solution is given in (\ref{sol1}--\ref{sol3}). For $\zeta_L=\zeta_S=1$
the roots are more complicated. If we restrict to $\epsilon_L=0$, we
find that they are
\begin{eqnarray}
\lambda&=&-{\bar\beta_n\over 2}\pm \sqrt{\left({\bar\beta_n\over
    2}\right)^2+\mu_L}, \br
-{\bar\beta_n\over 2}\pm \sqrt{\left({\bar\beta_n\over
    2}\right)^2+(2-\bar\rho)\mu_S}.
\end{eqnarray}
Diagonalisability of the scalar sector of ADM and NOR-A/B
with LS gauge is guaranteed in the generic case where
(with $\epsilon_L=0$)
\begin{eqnarray}
\label{shift1}
(2-\bar\rho)\mu_S-\mu_L&\ne&0, \\
\label{shift2}
|\mu_L-1|&>&|\bar\beta|, \\
\label{shift3}
|(2-\bar\rho)\mu_S-1|&>&|\bar\beta|
\end{eqnarray}
(Note that the equivalent conditions for the $\p_0$ case are obtained
by setting $\bar\beta^i$ to zero here.) However, with $ab=1$ condition
(\ref{shift2}) is not necessary, and (\ref{shift3}) is not necessary
for NOR with $s=1$.

In conclusion NOR-A/B with LS gauge can be made strongly hyperbolic
with $s=1$ for all values of $\beta^i$, at least for $\epsilon_L=0$.

Finally we note that the lapse driver alone, with fixed shift, is
strongly hyperbolic for NOR-B, but not for ADM (the vector sector
fails to be diagonalisable) or NOR-A (the scalar sector fails to be
diagonalisable. The same result was found in the ST system
\cite{SarbachTiglio}, which is equivalent to NOR with the lapse driver
and fixed shift.


\subsection{Symmetric hyperbolicity}


\subsubsection{NOR with live lapse and fixed shift}


The most general conserved energy has the form
\begin{equation}
\label{epsST}
\epsilon = \sum_{i=0}^5 c_i \epsilon_i,
\end{equation}
with
\begin{eqnarray}
\label{eps0}
\epsilon_0 &=& K_{ij}K^{ij}+\frac{1}{4}\g_{ij,k}{}^2 + d^i Y_i
-(ad^i+dt^i)X_i \\
&+& \frac{ab-1}{2}\,d^id_i+\frac{c-bd}{2}\,t^it_i+
\frac{1-2ab+\mu_L}{2}\,d^it_i, \nonumber\\
\epsilon_1 &=& K^2 + t^i \left[ Y_i - (a+3d)X_i
\right. \nonumber \\
&& \left. \qquad\quad +\frac{1}{4}(2-2ab+6c+6bd+\mu_L)t_i\right], \\
\epsilon_2 &=& \g^{ij}( d_i d_j - \g^{kn}\g^{lm}\g_{ik,l}\g_{jm,n}), \\
\epsilon_3 &=& X_i X^i , \\
\epsilon_4 &=& X_i Y^i , \\
\epsilon_5 &=& Y_i Y^i ,
\label{eps5}
\end{eqnarray}
where we have defined the shorthands
\begin{eqnarray}
X_i &\equiv& f_i-(b-\frac{\rho}{2})t_i+(b-1)d_i , \\
Y_i &\equiv& (\log Q)_{,i}+\frac{1}{2}(\sigma-\mu_L)t_i .
\end{eqnarray}
{ (This energy is found by writing down the most general $\epsilon$
and flux $\phi^i$ quadratic in $\g_{ij,k}$, $Q_{,i}$, $K_{ij}$ and
$f_i$ constructed using only these and $\g^{ij}$, and restricting
their 14+10 free coefficients by 19 conditions arising from energy
conservation $\dot\epsilon={\phi^i}_{,i}$ in the high-frequency
approximation. We do not give the fluxes here. Positive definiteness
of $\epsilon$ has not been imposed yet.)}  Energies $\epsilon_3$,
$\epsilon_4$ and $\epsilon_5$ are automatically conserved with zero
flux. Conservation of $c_0\epsilon_0+ c_1\epsilon_1$ requires
\begin{equation} \label{restriction}
c_0(1-2ab+\mu_L+2c-2bd)+2c_1(1-ab+3c-3bd) = 0,
\end{equation}
with non-zero flux, and conservation of $\epsilon_2$ (with non-zero
flux as well) poses no new condition on the parameters.  The
coefficient $c_0$ must be strictly positive and therefore there are
two possibilities: either $ab-1\ne 3(c-bd)$ and then $c_1$ is
determined by $c_0$ and the parameters $(a,b,c,d,\mu_L)$, or we have
the special case $ab-1=3(c-bd)=3(\mu_L-1)/4$ and then $c_0$ and $c_1$
are independent. Both NOR-A/B belong to that special case and require
$\mu_L=1$, which is harmonic slicing. Generic NOR however allows
for arbitrary values of $\mu_L$.

With the evolved lapse we obtain the same slicing as with fixed
densitised lapse if we set $\mu_L=\sigma$ and $\p_tQ=0$. Then $Y_i$
can be dropped from the energy and (\ref{epsST}--\ref{restriction})
reduces to the energy of NOR with densitised lapse and fixed shift
(\ref{fixedNORenergy}).

Fixing an overall factor by setting $c_0=1$, the energy has either 9
free objects (4 coefficients $c_i$ plus 5 parameters), or 8 free
objects in the special case (5 coefficients plus 3 parameters). The
ranges for those objects are restricted by the positivity conditions
on the energy.  The tensor and scalar sectors require
\begin{equation}
c_1 > -\frac{1}{3}, \qquad
-\frac{1}{4}<c_2<\frac{1}{2},
\end{equation}
and the vector sector requires positivity of a 4$\times$4 matrix,
which is difficult to convert into explicit inequalities for the
coefficients and parameters. Some necessary conditions are
\begin{equation}
c_3>0, \qquad c_5>0, \qquad c_4^2 < 4 c_3 c_5 .
\end{equation}
For example for NOR-A with $\mu_L=1$ and choosing $c_4=0$ we still
have 4 free coefficients, and a complete set of symmetric hyperbolicity
conditions are
\begin{eqnarray}
& c_1 > 0, \quad 0<c_2<\frac{1}{2}, 
\quad c_3>c_1+\frac{1}{4c_2}, & \nonumber  \\
& c_5 > \frac{(4c_1c_2+1)c_3}{4c_2(c_3-c_1)-1} . &
\end{eqnarray}


\subsubsection{NOR with live lapse and shift}


The most general conserved energy can be written as
\begin{equation}
\label{epsLS}
\epsilon = \sum_{i=0}^{10} c_i \epsilon_i,
\end{equation}
with $\epsilon_0,...,\epsilon_5$ as given before. Defining
\begin{eqnarray}
W_i &\equiv& \mu_S\left[s X_i +(bs-\frac{\bar\rho}{2})t_i
+(1-bs)\frac{t_i}{2}\right] +
\epsilon_S \left[Y_i + \frac{\mu_L}{2} t_i\right], \\
Z_i &\equiv& \mu_S\left[s X_i +(bs-\frac{\bar\rho}{2})\frac{t_i}{2}
\right] +
\epsilon_S \left[Y_i + \frac{\mu_L}{4} t_i\right],
\end{eqnarray}
we have five new energy terms containing derivatives of the shift vector
\begin{eqnarray}
\epsilon_6 &=& \bar\beta^i{}_{,j}\bar\beta^j{}_{,i} , \\
\epsilon_7 &=& \bar\beta^i{}_{,i} \bar\beta^j{}_{,j} 
- \bar\beta^i{}_{,j}\bar\beta^j{}_{,i} , \\
\epsilon_8 &=& \g^{ik}\g_{jl}\bar\beta^j{}_{,i}\bar\beta^l{}_{,k}- \bar\beta^i{}_{,j}\bar\beta^j{}_{,i} , \\
\epsilon_9 &=& K^i{}_j \bar\beta^j{}_{,i} -
\frac{\alpha}{2}W^id_i , \\
\epsilon_{10} &=& K \bar\beta^i{}_{,i} -
\frac{\alpha}{2}Z^it_i .
\end{eqnarray}
Assuming $b\ne 0$ and $\epsilon_L\ne \mu_L$ (two necessary conditions
for strong hyperbolicity in NOR), the coefficients $c_3, c_4, c_5$
are always determined by the rest of the coefficients:
\begin{eqnarray}
4 b c_3 &=& 2 a c_0+ 4s\mu_S c_8 +(2a+s\mu_S) c_9, \\
(\epsilon_L-\mu_L)c_4 &=& 2(a+d)c_0+2(a+3d)c_1+2s\mu_S c_6 \\
&& + (a+d+s\mu_S)c_9+(a+3d+s\mu_S)c_{10}, \nonumber \\
(\epsilon_L-\mu_L)c_5 &=& \epsilon_S c_6-(c_0+c_1)+\frac{\epsilon_S-1}{2}(c_9+c_{10}).
\end{eqnarray}
Energy conservation imposes 4 more conditions:
\begin{eqnarray}
\label{LScond1}
(ab-\mu_S)c_9 +4\mu_S(bs-1)c_8 &=& 0 , \\
\label{LScond2}
2\epsilon_L c_0  +
[2ab-1-2c+2bd-\epsilon_L\epsilon_S\nonumber \\
+\mu_S(\bar\rho-2)]c_9 
+2(ab-1-3c+3bd)c_{10}&&\nonumber \\+
4\mu_S(bs-1)c_6 &=& 0 , \\
\label{LScond3}
[2-2ab+6c-6bd-\epsilon_L\epsilon_S&&\nonumber \\
+\mu_L+\mu_S(\bar\rho-2)]c_{10}
\nonumber \\
+2[\mu_S(\bar\rho-2bs)-\epsilon_S\mu_L]c_6 && \nonumber \\
+2\epsilon_L c_1 +(1-2ab+2c-2bd+\mu_L)c_9 &=& 0 , \\
\label{LScond4}
c_0(1-2ab+\mu_L+2c-2bd)&& \nonumber \\+
2c_1(1-ab+3c-3bd)
+\mu_S(1-bs)c_{10} && \nonumber \\
 +
\frac{1}{2}[\mu_S(\bar\rho-2bs)-\epsilon_S\mu_L]c_9 &=& 0 .
\end{eqnarray}
For generic values of the parameters, these four conditions allow us
to solve for $c_6,c_8,c_9,c_{10}$, leaving only four free coefficients
$c_0,c_1,c_2,c_7$.  It is possible, however, to choose the
parameters so that some of those four conditions are
automatically obeyed, and then the number of free coefficients is
larger. For example, for NOR with implicit harmonic gauge and $s=1$
all four conditions are automatically obeyed, leaving eight free
coefficients $c_i$. Note that the condition (\ref{restriction}) has
now been transformed into (\ref{LScond4}). Note also that for
$\epsilon_L=0$ conditions (\ref{LScond1}, \ref{LScond2}, \ref{LScond3})
only involve the new coefficients $c_6,...,c_{10}$.

So far we have only looked at energy conservation. Symmetric
hyperbolicity requires positivity of the conserved energy. Ideally, we
would like to translate positivity into two sets of inequalities:
ranges of evolution parameters that lead to a symmetric hyperbolic
system, followed by (parameter-dependent) inequalities for the
free coefficients of the conserved energy given above. We achieved
this for NOR with fixed gauge in \cite{bssn2}, but have not managed 
for the systems studied here.

Sometimes, for example for the LS system \cite{LindblomScheel}, it is
possible to choose the coefficients of a generic energy to make it
positive, and then solve the energy conservation equations for the
evolution parameters of the system. This method proves that the
system is symmetric hyperbolic for some values of the parameters, but
works only if there are enough parameters in the evolution
system. It may also be difficult to find energies relating to
specific interesting values of the evolution parameters.

For NOR with LS gauge, we only have 12 parameters, namely $(a,b,c,d)$,
$(\mu_L,\mu_S,\epsilon_L, \epsilon_S)$ and $(\rho,\bar\rho,\sigma,s)$,
to solve 15 linear equations imposed by energy conservation on 19
coefficients of a general energy.
Of the 19 coefficients, 11 multiply squares (diagonal terms) and 8
multiply mixed products (off-diagonal terms). Setting 7 of the
off-diagonal terms equal to zero, we have been able to solve the
remaining positivity conditions and all energy conservation equations
to obtain a 9-parameter (7 evolution parameters and 2 energy
coefficients) family of symmetric hyperbolic systems.
Three of the five relations among the parameters are
$\epsilon_L=0$, $bs=1$ and $\mu_L\epsilon_S=(\bar\rho-2)\mu_S$; the
other two are rather longer.
However, there
is no reason to assume that any of the positivity conditions fail if
all off-diagonal coefficients of the energy are given sufficiently
small non-zero values. Therefore we believe that our 9-parameter
family is embedded in a 16-parameter open set of symmetric hyperbolic
formulations, which could be expressed in terms of the 12 evolution
parameters and 4 energy coefficients.

In particular, we have been able to construct positive and conserved
energies for relevant choices of the parameters: for example for NOR-B
with $\mu_L=\mu_S=s=1$,
$\epsilon_L=0$, $\epsilon_S=-1$,  the four conservation equations
reduce to $c_9(\bar\rho-1)=0$ and
$(2c_6+c_{10})(\bar\rho-1)=0$. For $\bar\rho=1$ the eight coefficients
are free and a possible positive energy is given by
\begin{eqnarray}
c_0=1, \quad c_1=112, \quad c_2=-\frac{3}{16}, \quad c_6=12, \\
c_7=-5 \quad c_8=16, \quad c_9=-8, \quad c_{10}=-8.
\end{eqnarray}
We can also construct a positive energy for BSSN-C
($\rho=\bar\rho=2/3$, $\sigma=0$) choosing 6 free coefficients,
although we have only been to solve the inequalities by finding
specific sets of energy coefficients, rather than by giving ranges.


\section{Hyperbolicity with implicit hyperbolic lapse and explicit
  hyperbolic shift drivers}
\label{section:puncturegauge}


Recently, Campanelli {\it et al.} \cite{Campanelli} and Baker {\it et al.}
\cite{Baker}, followed by Diener {\it et al.} \cite{Diener} {\it et al.} and
Herrmann {\it et al.} \cite{Herrmann} have reported significant progress in
binary black hole evolutions with the BSSN-C formulation using an
implicit hyperbolic lapse driver combined with an explicit hyperbolic
shift driver. The gauge choice differs slightly between these four
groups. We shall look at a family that includes all four cases plus a
fifth that we suggest as a simpler alternative. In the following we
look at BSSN-C in the form of NOR-B ($a=b=1$, $c=d=-1/3$) with the
further choice $s=1$, $\rho=\bar\rho=2/3$, so that
$f^i=\g^{1/3}\tG^i$. We also set $\sigma=0$. The gauge choice we
consider is
\begin{eqnarray}
\left(\p_0+\zeta_L\p_{\bar\beta}\right)\ln\alpha&\simeq &-\mu_L K, \\
\left(\p_0+\zeta_S\p_{\bar\beta}\right)\beta^i &\simeq &b^i, \\
\left(\p_0+\zeta_b\p_{\bar\beta}\right)b^i
&\simeq&\mu_S\alpha\left(\p_0+\zeta_f\p_{\bar\beta}\right)\bar f^i.
\end{eqnarray}
(In each case, the factor of proportionality between our variable
$b^i$ and the variable $B^i$ in which these gauges are originally
formulated is different. We have introduced $b^i$ to simplify the
comparison of the principal terms.) In the principal part, this
corresponds to the gauge choices of the four groups with the values of
the six parameters given in 
Table~\ref{table:punctures}. All four groups set $\zeta_S=\zeta_b=1$.
By eliminating the variable $b^i$ in favour of a second time
derivative of $\beta^i$, and noting that $b^i$ does not appear
anywhere else in the evolution equations, we see that
$(\beta_S,\beta_b)=(0,1)$ and $(1,0)$ would be equivalent
in the principal part.

\begin{table*}
\begin{tabular}{|l|cccccc|l|l|}
\hline
& $\zeta_L$ & $\zeta_S$ & $\zeta_b$ &$\zeta_f$ & $\mu_L$ & $\mu_S$ & 
speeds & strongly hyperbolic for \\
\hline
Campanelli {\it et al.} & 0 & 1 & 1 & 1 & ${2\over\alpha}$ & 
${3\over
 4}\alpha^{-2}\g^{1/3}$ &
$(0,\pm
 1,\pm\sqrt{\mu_L},-\bar\beta_n,\lambda_{2\pm},
\pm\sqrt{\mu_S}$ &
$|\bar\beta|<|(3\mu_L-4\mu_S)/(3\sqrt{\mu_L})|$, \
$|\bar\beta|<\sqrt{\mu_S}$
\\
Baker {\it et al.} & 0 & 1 & 1 & 0 & 
${2\over\alpha}$ & ${3\over 4}\alpha^{-1}\g^{1/3}$ & 
$(0,\pm 1,\pm\sqrt{\mu_L},\lambda_{3\pm},\lambda_{4\pm})$ &
$|\bar\beta|<\sqrt{4\mu_S/3}$, \
$|\bar\beta|<|\sqrt{\mu_L}-\sqrt{4\mu_S/3}|$\\
Diener {\it et al.} & 1 & 1 & 1 & 1 & 
${2\psi_{\rm BL}^m\over\alpha}$ & 
${3\over 4}{\alpha^{p-2}\over \psi_{\rm BL}^n}\g^{1/3}$ & 
$(0,\pm 1,-\bar\beta_n,\lambda_{1\pm},\lambda_{2\pm},
\pm\sqrt{\mu_S})$ &
$\mu_L\ne 4\mu_S/3$, \ $|\bar\beta|<\sqrt{\mu_S}$
  \\
Herrmann {\it et al.} & 1 & 1 & 1 & 0 & 
${2\over\alpha}$ & ${3\over 4}\alpha^{-2}\g^{1/3}$ & 
$(0,\pm 1,-\bar\beta_n,
\lambda_{1\pm},\lambda_{3\pm},\lambda_{4\pm})$ &
$|\bar\beta|<|(3\mu_L-4\mu_S)/(2\sqrt{3\mu_S})|$ \\
Beyer-Sarbach & 0 & 0 & 0 & 0 & $f$ & $GH$ & 
$(0,\pm 1,\pm\sqrt{\mu_L},\pm \sqrt{\mu_S},\pm \sqrt{4\mu_S/3})$ &
$\mu_S\ne \mu_L$, $\mu_S\ne 3\mu_L/4$ \\
\hline
\end{tabular}
\caption{ Parameter values, characteristic speeds, and conditions for
strong hyperbolicity for the four ``puncture evolution'' codes { and the
variant of Beyer and Sarbach. For comparison with these authors, we
have expressed our parameters $\mu_L$ and $\mu_S$ in their notation,
but for comparison between the gauges we give the characteristic
speeds in our notation.}
\label{table:punctures}
}
\end{table*}

With the shorthands
\begin{eqnarray}
\lambda_{1\pm}&\equiv&-{\bar\beta_n\over 2}\pm \sqrt{\left({\bar\beta_n\over
    2}\right)^2+\mu_L}, \\
\lambda_{2\pm}&\equiv&-{\bar\beta_n\over 2}\pm \sqrt{\left({\bar\beta_n\over
    2}\right)^2+{4\over 3}\mu_S}, \\
\lambda_{3\pm}&\equiv&-\bar\beta_n\pm \sqrt{{4\over 3}\mu_S}, \\
\lambda_{4\pm}&\equiv&-{\bar\beta_n\over 2}\pm \sqrt{\left({\bar\beta_n\over
    2}\right)^2+\mu_S},
\end{eqnarray}
the second column of Table~\ref{table:punctures} then gives all
characteristic speeds of these four systems. In all four systems
strong hyperbolicity breaks down when some of the speeds coincide. The
conditions for avoiding this are listed in the third column of the
table. It is possible that these inequalities hold for the specific
binary black hole evolutions because $\bar\beta^i$ was sufficiently
small in them, but one would expect numerical difficulties to appear
for $\bar\beta^i$ sufficiently large to violate one of these
conditions, for example in co-rotating coordinates. It would be simple
and interesting to verify this.

If we set $\zeta_L=\zeta_S=\zeta_b=\zeta_f=0$, all speeds become independent
of $\bar\beta^i$. They are
\begin{equation}
(0,\pm 1,\pm\sqrt{\mu_L},\pm \sqrt{\mu_S},\pm \sqrt{4\mu_S/3})
\end{equation}
and the conditions for strong hyperbolicity are $\mu_S\ne \mu_L$ and
$\mu_S\ne 3\mu_L/4$, which can now easily be arranged through the
choice of $\mu_L$ and $\mu_S$ as functions of $\alpha$, for example
$\mu_S=3/4$ and $\mu_L=2/\alpha$, which would not coincide as long as
$\alpha<2$, independently of the shift. { Strong hyperbolicity of
BSSN with this gauge choice was shown in
\cite{BeyerSarbach}. Hyperbolicity of the Z4 formulation with a family
of gauges with the same principal part and, in our notation,
$\zeta_f=\zeta_S=0$ and $\zeta_L$ and $\zeta_b$ free, was examined in
\cite{BonaLehnerPalenzuela}.}


\section{Mode analysis results}
\label{section:modeanalysis}


\subsection{Formalism}


If the complete evolution system is not hyperbolic, we can use mode
analysis to obtain at least necessary conditions for
well-posedness. We make two approximations: we linearise around a
background, and we approximate all background-dependent coefficients
of the linearised equations as constant (``frozen''). Physically this
corresponds to investigating small amplitude, high frequency
perturbations. { (The results obtained will in general depend both on
the background solution, and the choice of gauge on it.)}

In the resulting linear problem with constant coefficients we
take a Fourier transform
\begin{equation}
u(x^i,t)\equiv \int e^{i\omega_ix^i}\hat u(\omega_i,t) \,d^3\omega
\end{equation}
and consider the evolution of one Fourier mode at a
time. Well-posedness of the Cauchy problem means that
$||u(\cdot,t)||\le f(t) ||u(\cdot,0)||$ with $f(t)$ independent of
the initial data $u(x^i,0)$. This is the case if and only if a similar
bound can be obtained for $\hat u(\omega_i,t)$ with $f(t)$ independent
of $\omega_i$. To establish well-posedness it is sufficient to
consider only certain leading order terms. This is well-known for
strongly hyperbolic first-order systems \cite{GKO}, but we need a
generalisation which is given in the following

{\bf Theorem~1:} {\em Consider a linear problem with constant coefficients
which in pseudo-differential form is
\begin{eqnarray}
\p_t\hat u(\omega_i,t)&=&[M(\omega_i)+M'(\omega_i)]\hat u(\omega_i,t), \\
\hat u(\omega_i,0)&=&\hat u_0(\omega_i).
\end{eqnarray}
The initial-value problem is well-posed if $M(\omega_i)$ is diagonalisable,
\begin{equation}
M(\omega_i)=T^{-1}(\omega_i)\Lambda(\omega_i)T(\omega_i)
\end{equation}
with $\Lambda(\omega_i)$ diagonal, and if $M(\omega_i)$ and $M'(\omega_i)$ obey the
bounds
\begin{eqnarray}
|T(\omega_i)|\,|T^{-1}(\omega_i)|&\le& K_1, \\
|{\rm Re} \Lambda(\omega_i)| &\le& K_2, \\
|M'(\omega_i)|&\le& K_3,
\end{eqnarray}
with $K_1$, $K_2$ and $K_3$ independent of $\omega_i$.}

{ For a more general statement in the context of semigroup theory see
\cite{Beyerlectures}. An elementary proof of our version of the theorem}
is given in Appendix~\ref{appendix:proof}, and a 
technical complication in applying the theorem to second-order in
space systems is resolved in Appendix~\ref{appendix:reduction}.

Clearly the well-posedness of the linearised, frozen-coefficients
problem is necessary for the well-posedness of the full problem. For
strongly hyperbolic problems well-posedness of the linearised
frozen-coefficient problem around any background, together with
smoothness of $T(\omega_i)$, is also sufficient \cite{GKO}, but we do
not know if that is true for the wider class of problems considered
here.


\subsection{Implicit hyperbolic drivers}


To establish notation we recast the problem of strong hyperbolicity of
the NOR formulation with implicit hyperbolic lapse and shift drivers,
which was already discussed in Section~\ref{section:hyperbolicity},
into the language of pseudo-differential operators. We define
\begin{eqnarray}
\hat u &\equiv& (i\omega\,\hat v,\hat w), \\
\hat Z&\equiv&i\omega\hat z,
\end{eqnarray}
where $\omega\equiv\sqrt{\g^{ij}\omega_i\omega_j}$ and $z=(\ln\alpha,\beta^i)$.
The factors $i\omega$ appear because this is a pseudo-differential
reduction to first order of a second-order in space system
\cite{reduct}. We define the pseudo-differential equivalent of the
derivative operator $\p_0$,
\begin{equation}
\hat\p_0\equiv \alpha^{-1}(\p_t-i\omega \beta_n I).
\end{equation}
The evolution of the gauge variables has the principal part $\dot
z\simeq w+\p v+\p z$, and combining this with (\ref{vdot},\ref{wdot})
the coupled pseudo-differential equations are
\begin{eqnarray}
\hat\p_0\left(\begin{array}{c}\hat u\\ \hat Z\end{array}\right)
&\simeq& i\omega  \left(\begin{array}{cc} P & Q \\
V & W \end{array}\right)
\left(\begin{array}{c}\hat u\\ \hat
    Z\end{array}\right),
\end{eqnarray}
where $P$, $Q$, $V$ and $W$ are independent of $\omega$ but depend on
$n_i\equiv \omega_i/\omega$. Here $\simeq$ means that terms
corresponding to the lower-order part $M'(\omega_i)$ have been
neglected. (In the context of a pseudo-differential reduction to first
order there is a slight complication in the definition of
$M'(\omega_i)$, which is discussed in
Appendix~\ref{appendix:reduction}.) $P$ and $Q$ can be read off from
(\ref{gdot}-\ref{fdot}). The nontrivial part of the matrix $V$ is
given by (\ref{fbardef}). $W$ contains the parameters $\zeta_L$,
$\zeta_S$, $\epsilon_L$ and $\epsilon_S$ defined in
(\ref{LS1}-\ref{LS2}).

We have analysed this system above in
Sect.~\ref{section:hyperbolicity}, and do not discuss it further here.


\subsection{Explicit parabolic drivers}


These have the general form $\dot z\simeq \dot w+ \p\dot v+\p\p z$. The
pseudo-differential form of the coupled equations is
\begin{eqnarray}
\hat \p_0
\left(\begin{array}{c}\hat u\\ \hat Z\end{array}\right)
&\simeq& 
\left(\begin{array}{cc} i\omega P & i\omega Q \\
-\omega^2 R+i\omega R' & -\omega^2 S+i\omega S' \end{array}\right) 
 \left(\begin{array}{c}\hat u\\ \hat
    Z\end{array}\right),\br
\label{EP1}
\end{eqnarray}
where the matrices $R$, $S$, $R'$ and $S'$ are independent of
$\omega$. Note that in order to apply Thm.~1 we need to work with the
lower-order terms $R'$ and $S'$. To calculate them we need to
explicitly linearise the gauge conditions.  If the principal part
matrix defined in (\ref{EP1}) obeys the conditions of Thm.~1, we
obtain a necessary condition for well-posedness in $L^2(w,\p v, \p
z)$. Alternatively, we can use a different pseudo-differential
reduction of the form
\begin{eqnarray}
\hat \p_0
\left(\begin{array}{c}\hat u\\ \hat z\end{array}\right)
&\simeq& 
\left(\begin{array}{cc} i\omega P & -\omega^2 Q +i\omega Q' \\
i\omega R& -\omega^2 S+i\omega S' \end{array}\right) 
 \left(\begin{array}{c}\hat u\\ \hat
    z\end{array}\right),\br
\label{EP2}
\end{eqnarray}
The conditions for Thm.~1 are then a necessary condition for
well-posedness in $L^2(w,\p v, z)$. The lower-order terms that must be
kept are now $S'$ and $Q'$.

When either set of lower-order terms is kept, the resulting principal
part matrix $M(\omega_i)$ becomes very complicated. In particular, the
eigenvalues depend on the background spacetime, and the matrix no
longer splits into block-diagonal form with scalar, vector and tensor
blocks. In order to make progress, we linearise around Minkowski
spacetime using Cartesian coordinates. 

In the Cartesian Minkowski
case the lower order terms $R'$, $S'$ and $Q'$ all vanish, and
we are left with the simpler matrices $M(\omega_i)$ of the form
\begin{equation}
\left(\begin{array}{cc} i\omega P & i\omega Q \\
-\omega^2 R& -\omega^2 S \end{array}\right)
\hbox{ or }
\left(\begin{array}{cc} i\omega P & -\omega^2 Q \\
i\omega R& -\omega^2 S \end{array}\right),
\end{equation}
which are similar, and therefore equivalent for the purposes of
Thm.~1. The diagonalisability and eigenvalues of these matrices
already give interesting necessary conditions for well-posedness.

We consider explicit parabolic
drivers in the form
\begin{eqnarray}
\p_0\ln\alpha &\simeq& -\nu_L \p_0 K, \label{p0alpha}\\
\label{foo}
\p_0 \beta^i &\simeq& -\nu_S \alpha \p_0 \bar f^i.
\end{eqnarray}
Note that when only principal terms are considered we can pull
the factor of $\alpha$ out of the time derivative in (\ref{foo}). $R$
and $S$ can then be read off from (\ref{Kdotgaugepp}) and
(\ref{fbardotpp}). We find that the simplified necessary conditions
based on linearisation around Minkowski are obeyed for this gauge with
ADM and NOR-B, but not NOR-A (the scalar sector fails to be
diagonalisable), for $\bar\rho<2$, $\nu_L>0$, $\nu_S>0$ and $\nu_L\ne
(2-\bar\rho) \nu_S$. The eigenvalues are $0$, $\pm i\omega$,
$-\nu_L\omega^2$, and $-(2-\bar\rho)\nu_S\omega^2$.

The lapse condition (\ref{p0alpha}) with {\em fixed} shift
together with ADM or NOR-A/B is ill-posed according to Thm.~1.  This can
be understood roughly as follows: The principal part of the lapse
evolution equation is $\dot\alpha\sim\Delta\alpha$, which is
autonomous. $\alpha$ can therefore be considered as a a given function
in the principal part of the remaining evolution equations, and so
their principal part is the same as for fixed (undensitised) lapse
(and fixed shift).

To investigate the explicit parabolic drivers using $\p_t$ on
both the left and right-hand sides, that is,
(\ref{dotalphadotK},\ref{parabolicGammadriver}), we consider the
matrix
\begin{equation}
\left(\begin{array}{cc} i\omega P & i\omega Q \\
-\omega^2 R + i\omega \bar \beta_n T & -\omega^2 S -i\omega \bar
\beta_n I\end{array}\right)
\end{equation}
(Again we drop the lower order terms by linearising around Minkowski.)
Neither ADM or NOR-A/B are diagonalisable with this gauge, except of
course for $\bar \beta_n=0$. 

Because the explicit parabolic lapse driver with fixed shift
using $\p_0$ alone is not well-posed (as we saw above), neither is the
version using $\p_t$ as the two coincide for $\bar\beta_n=0$.


\subsection{Implicit parabolic drivers} 


These have the general form $z\simeq w+\p v$. After linearisation, in
pseudo-differential form
\begin{equation}
\hat Z\simeq (i\omega T+T')\hat u,
\end{equation}
where $T$ and $T'$ are independent of $\omega$. In order to calculate
the lower order term $T'$, we again need to explicitly linearise the
gauge conditions.  Eliminating $\hat Z$, we find
\begin{eqnarray}
\hat \p_0\hat u\simeq \left[-\omega^2 QT+i\omega (P+QT')\right] \hat u.
\end{eqnarray}
We can again make progress by linearising around Minkowski spacetime
in standard coordinates, so that $T'$ vanishes. $T$ is equal to
$V$ with $\mu_L$ and $\mu_S$ replaced by $\nu_L$ and $\nu_S$.

The conditions of Thm.~1 are obeyed for the implicit parabolic
drivers (\ref{implicitparaboliclapse}, \ref{implicitparabolicshift})
with ADM and NOR, with the same inequality conditions as for
the explicit parabolic drivers. The eigenvalues are also the same as
for the explicit parabolic drivers. 

The implicit parabolic lapse driver (\ref{implicitparaboliclapse})
with {\em fixed} shift obeys the conditions of Thm.~1 for NOR-B for
$\nu_L>0$.  The eigenvalues are $0$, $\pm i\omega$ and
$-\nu_L\omega^2$. This is not true for NOR-A (because the scalar
sector is not diagonalisable) or ADM (because the vector sector is not
diagonalisable).


\subsection{Elliptic freezing conditions}


These have the general form $\dot w+\p\dot v\simeq 0$. After linearisation,
in pseudo-differential form, 
\begin{equation}
0\simeq(-\omega^2 R+i\omega R')\hat u+(-\omega^2S+i\omega
S')\hat Z,
\end{equation}
where $R'$, $R$, $S'$ and $S$ can be chosen to be the same
matrices as above. Eliminating $\hat Z$, we obtain the reduced system
\begin{eqnarray}
\hat \p_0 \hat u\simeq i\omega (P-QS^{-1}R) \hat u.
\end{eqnarray}
Note that this algebraic reduction can only be carried out in
pseudo-differential form. Intuitively speaking, $R'$ and $S'$ do
not appear in the principal part because of the smoothing effect of
solving an elliptic equation. 

The elliptic freezing conditions $\p_0 K=0$, $\p_0 \bar f^i=0$ obey
the conditions of Thm.~1 together with ADM and NOR-B but not NOR-A
(the scalar sector is not diagonalisable) for $\bar\rho\ne 2$. The
speeds are $0$ and $\pm 1$. 

The elliptic lapse condition alone, with fixed shift, is not
diagonalisable for either of the three systems. This slightly
surprising result can be explained as follows: The only component of
$\p_0 K_{ij}\simeq -(\ln\alpha)_{,ij}+\dots$ in the
pseudo-differential approach is $\hat \p_0 \hat K_{nn}\simeq \omega^2
\hat {(\ln \alpha)}+\dots$. But the elliptic equation for $\alpha$ is
$-\omega^2 \hat {(\ln \alpha)}\simeq 0$. The scalar sector of the
principal part is then the same as for fixed (undensitised) lapse, and so
is not diagonalisable.

However, if one formally enforces the constraint $K=\phi(x)$ by
setting $\hat K_{qq}\simeq -\hat K_{nn}$ the scalar sector of NOR-A/B
(but not ADM) becomes diagonalisable, and the resulting system obeys
the conditions of Thm.~1. In the BSSN formulation $K$ is an explicit
variable, and one can enforce $K=\phi(x)$ by not evolving this
variable when BSSN is used with maximal slicing and fixed shift
\cite{AlcubierreBSSN}. With this modification the system is
well-posed, but if $K$ was evolved it would be ill-posed.


\subsection{Implicit hyperbolic drivers with parabolic damping
  terms}


The gauge conditions (\ref{Shibatalapse},\ref{Shibatashift}),
linearised around Minkowski in Cartesian coordinates, give
\begin{eqnarray}
\hat\p_0\left(\begin{array}{c}\hat u\\ \hat Z\end{array}\right)
&\simeq&   \left(\begin{array}{cc} i\omega P & i\omega Q \\
i\omega V -2\omega^2 R & i\omega W -2\omega^2 S\end{array}\right)
\left(\begin{array}{c}\hat u\\ \hat
    Z\end{array}\right), \br
\end{eqnarray}
For simplicity we restrict the hyperbolic gauge conditions to
$\zeta=\epsilon_L=\epsilon_S=0$, and we consider ADM. Without the
damping terms, that is with $\nu_L=\nu_S=0$, the system is strongly
hyperbolic for $\mu_L>0$, $\mu_S>0$ and $\bar\rho< 2$, with
eigenvalues $0$, $\pm i\omega$, $\pm i\omega \sqrt{\mu_L}$, $\pm
i\omega\sqrt{(2-\bar\rho)\mu_S}$ and $\pm i\omega \sqrt{\mu_S}$,
provided that $\mu_L\ne 1$, $\mu_S\ne 1/(2-\bar\rho)$ and $\mu_S\ne
\mu_L/(2-\bar\rho)$, corresponding to real speeds. With the damping
turned on, the pair of eigenvalues $\pm i\omega \sqrt{\mu_L}$ becomes
\begin{eqnarray}
-\omega^2 \sqrt{\nu_L}\pm\sqrt{\omega^4 \nu_L^2-\omega^2\mu_L},
\end{eqnarray}
and similarly for the other eigenvalues. This has negative real part
for all $\omega>0$.  However, the scalar sector fails to be
diagonalisable if these two eigenvalues coincide, that is for
$\omega=\sqrt{\mu_L}/\nu_L$. Similar problems arise for
$\omega=\sqrt{\mu_L}/(\sqrt{2-\bar\rho}\nu_L)$ in the scalar sector
and $\omega=\sqrt{\mu_L}/\nu_L$ in the vector sector.  In numerical
applications, choosing $\nu_S,\nu_L\propto \Delta x$ with a
sufficiently small constant of proportionality makes the problematic
wave numbers larger than the grid frequency, and this ill-posedness
should then not lead to numerical problems \cite{Shibata}.
Similar results hold for NOR-A/B at least in the special cases $s=0$ or
$\rho=2$.


\section{Conclusions}
\label{section:conclusions}


Continuing our research programme on
second-order in space, first-order in time formulations of the
Einstein equations \cite{bssn1,bssn2,reduct},
we have reviewed the relationship between the BSSN formulation of the
Einstein equations that is widely used in numerical relativity, and
the NOR formulation (NOR-A) that was suggested as a simpler
alternative to BSSN. We have showed that a variant of NOR, (NOR-B)
has the same principal part as BSSN when the latter is formally
restricted to solutions of the algebraic constraints (BSSN-C).

We have shown that the principal parts of the KST formulation and the
NOR formulation are equivalent when the gauge is fixed. Each
formulation has five parameters, and we have given four relations
between the two sets of parameters. One parameter in each formulation
does not have a counterpart in the other formulation and does not
influence the level of hyperbolicity. The analogy also holds with an
evolved lapse, and symmetric hyperbolic cases exist for both fixed
gauge and live lapse. With an evolved lapse and shift, KST has
four parameters more than NOR with the same gauge.

We have reviewed various differential equations for the lapse and
shift currently in use as coordinate conditions in numerical
relativity, with an emphasis of their origin as symmetry-seeking
coordinates, and their classification as elliptic, parabolic,
hyperbolic, or neither. These are all based on $K$-freezing shift and
$\Gamma$-freezing lapse, which in turn are related to the well-known
maximal slicing, minimal distortion gauge. 

From the point of view of ``symmetry-seeking'' gauge conditions it is
more natural to have evolution equations for the gauge of the form
$\p_t\alpha,\p_t\beta=\dots$, while the ADM evolution equations are
naturally of the form $\p_0\g_{ij},\p_0 K_{ij}=\dots$. With this
mixture, there are two sets of light cones in the system, one centred
around $n^a$ and one around $(\p/\p t)^a$. Where these overlap, there
is a danger of strong hyperbolicity breaking down, but we have shown
that some $\p_t$ gauges can be implemented in $\p_0$ form, which would
avoid this problem.

We have analysed the hyperbolicity of the ADM, NOR-A and NOR-B
formulations with the most general implicit hyperbolic lapse and shift
drivers. Interestingly, NOR-A and NOR-B can be made strongly
hyperbolic even when using some live gauges with $\p_t$. 

We have also
investigated the hyperbolicity of NOR-B (equivalent to BSSN-C) with a
family of implicit hyperbolic lapse and explicit hyperbolic shift
drivers that includes the gauges used by four different research
groups in ``moving puncture'' evolutions. We find that all these
gauges become ill-posed for large enough values of the shift, and
propose a simple modification (replacing $\p_t$ by $\p_0$) that is
strongly hyperbolic for arbitrary shift.

For certain families of elliptic gauge conditions, parabolic drivers,
and hyperbolic drivers with a heat equation-type damping, where the
notion of hyperbolicity does not apply, we have carried out the mode
analysis and have checked a necessary condition for well-posedness of
the Cauchy problem.

Khoklov and Novikov \cite{KN} have investigated the
well-posedness of a number of gauges {\em independently} of any
particular formulation in the frozen coefficient approximation. Their
well-posedness is therefore a necessary condition for
ours. We review their method and compare results in
Appendix~\ref{appendix:KN}.

{Our results for puncture gauges are summarised in
Table~\ref{table:punctures} and our results for all other gauges} in
Table~\ref{table:overview}. They are positive for most combinations of
gauge and formulation, with a few interesting exceptions. They provide
a theoretical underpinning to some conditions already in use in
numerical relativity, and suggest certain improvements.

All equations in section \ref{section:hyperbolicity} were derived
using {\em xTensor}, an
open-source {\em Mathematica} package for abstract tensor calculations,
developed by J.M.M. It is available under the GNU Public License from
http://metric.iem.csic.es/Martin-Garcia/xAct.

\begin{table}
\begin{tabular}{|p{3cm}|c|c|}
\hline
type & lapse + shift & lapse only \\
\hline
impl. hyp. using $\p_0$ & ADM, NOR-A/B & NOR-A/B \\
impl. hyp. using $\p_t$ & NOR-A/B & NOR-B \\
\hline
expl. par. using $\p_0$ & ADM, NOR-B  & no \\
expl. par. using $\p_t$ & no & no \\
implicit parabolic & ADM, NOR-A/B  & NOR-B  \\
elliptic & ADM, NOR-B  & no \\
impl. hyp. using $\p_0$ with par. damping & ADM, NOR-A/B  & ? \\
\hline
\end{tabular}
\caption{Overview of well-posedness results for combinations of gauge
  conditions and formulations of the Einstein equations. In the upper
  half of the table, strong hyperbolicity has been checked. In the lower
  half of the table only only a necessary condition (Thm. 1) has been
  checked. A question mark indicates that we have not done
  the calculation.
\label{table:overview}
}
\end{table}

\acknowledgments

We would like to thank G. Calabrese, L. Lehner, L. Lindblom and
O. Sarbach for helpful discussions. JMM was supported by the
Spanish MEC under the research project FIS2005-05736-C03-02.



\appendix


\section{Proof of Theorem~1}
\label{appendix:proof}


We go to the diagonal basis $\hat U=T\hat u$. It obeys
\begin{equation}
\p_t \hat U=(\Lambda+TM'T^{-1})\hat U,
\end{equation}
and so its norm squared obeys
\begin{equation}
\p_t (\hat U^\dagger \hat U)\le 2(K_2 + K_1 K_3)\hat U^\dagger \hat U.
\end{equation}
This can be integrated to give
\begin{equation}
|\hat U(\omega_i,t)|\le e^{(K_2+K_1K_3)t}|\hat U(\omega_i,0)|,
\end{equation}
and going back to the original basis we have
\begin{equation}
|\hat u(\omega_i,t)|\le K_1 e^{(K_2+K_1K_3)t}|\hat u(\omega_i,0)|,
\end{equation}
that is, the growth is bounded independently of the wavenumber
$\omega$. Taking the inverse Fourier transform and using Parseval's
Theorem, we obtain the $L^2$ estimate
\begin{equation}
||u(\cdot,t)||\le K_1 e^{(K_2+K_1K_3)t} ||u(\cdot,0)||
\end{equation}
in real space.

The well-known result that lower-order terms do not affect the
well-posedness of strongly hyperbolic systems (see for example Theorem
4.3.2 of \cite{GKO}) is a special case of Thm.~1. A system of linear
first-order evolution equations is strongly hyperbolic if
$M(\omega_i)=i\omega_i P^i\equiv i\omega P_n$, with $\omega_i\equiv
\omega n_i$ and $P_n\equiv n_i P^i$, and where $P_n$ is
diagonalisable with real eigenvalues for every $n_i$ (so that
${\rm Re}\Lambda=0$), $T$ depends smoothly on $n_i$ but not on $\omega$
(and so is bounded), and $M'$ is independent of $\omega_i$ (and so is
bounded).


\section{Pseudo-differential reduction to first order}
\label{appendix:reduction}


The pseudo-differential form of the linearisation of
(\ref{vdot},\ref{wdot}) on a constant background, neglecting the terms
in $z$ and writing all (linearised) lower-order terms, is
\begin{eqnarray}
\hat\p_0\left(\begin{array}{c}\hat v\\ \hat w\end{array}\right)
&\simeq&   \left(\begin{array}{cc} i\omega A+A' & B \\
-\omega^2 C +i\omega C' +C'' & i\omega D + D'\end{array}\right)
\left(\begin{array}{c}\hat v\\ \hat
    w\end{array}\right), \br
\end{eqnarray}
where the matrices $A$, $A'$ etc. are independent of $\omega$ but
depend on $n_i$. Replacing $\hat v$ by $\hat V\equiv i\omega \hat v$
multiplies the first row by $i\omega$ and divides the first column by
$i\omega$, so that the highest power of $\omega$ is the first
power. This constitutes a pseudo-differential reduction to first
order.

The terms proportional to $i\omega$ form $M(\omega_i)$ and the terms
of $O(1)$ go into $M'(\omega_i)$. The term $(i\omega)^{-1}C''$ poses a
problem, as it is not bounded independently of $\omega_i$. Intuitively,
it should be part of $M'(\omega_i)$ because we are interested mainly
in the limit $\omega\to\infty$. We can obtain a rigorous result to the
same effect if we retain $\hat v$ in the system. We then have
\begin{eqnarray}
\hat\p_0\left(\begin{array}{c}\hat v\\ \hat V \\\hat
w\end{array}\right) 
&\simeq& i\omega \left(\begin{array}{ccc} 0 & 0 & 0 \\ 0 &
A & B \\ 0 & C & D \end{array}\right) \left(\begin{array}{c}\hat v\\
\hat V \\ \hat w\end{array}\right) \br
+ \left(\begin{array}{ccc} A' & A & B \\ 0 &
A' & 0 \\ C'' & C' & D' \end{array}\right) \left(\begin{array}{c}\hat v\\
\hat V \\ \hat w\end{array}\right)
\end{eqnarray}
For the purposes of Thm.~1 we only need to analyse $M(\omega_i)$, which
here only means the block $((A,B),(C,D))$, which is called $P$ in the
main text. 


\section{Analysis of pure gauge perturbations}
\label{appendix:KN}


The effect of a set of gauge conditions can be evaluated in purely
geometric terms, independently of any formulation of the Einstein
equations. In this approach, one considers a fixed spacetime obeying
all 10 Einstein equations. One evolves only the four coordinates
$x^\mu$ and the vector $(\p/\p t)^a$ on this spacetime background,
treating $x^\mu$ as four scalar fields and $(\p/\p t)^a$ as a vector
field on the background spacetime. To make connection with the 3+1
split, let $n^a$ be the unit normal on the surfaces of constant $x^0$,
and let $(\p/\p t)^a=\alpha n^a +\beta^a$, where $\beta^a n_a\equiv
0$. The resulting non-linear evolution system for
$(x^\mu,\alpha,\beta^i)$ and a numerical
implementation are described in \cite{thornexact}.

A linearisation of this scheme is used in \cite{KN} to investigate the
well-posedness of gauges. The gauge perturbation is parameterised by a
vector field $\psi_a$, which gives rise to a perturbation of the
4-metric $\delta g_{ab}=-2\nabla_{(a}\psi_{b)}$. The 10 components of
this equation can be written in a 3+1 split as
\begin{eqnarray}
\dot\psi_0&=&\alpha
\delta\alpha-\beta^i\delta\beta_i-\beta^i\beta^j\psi_{,ij} \br
+\left({\Gamma^\mu}_{00}+\beta^i\beta^j{\Gamma^\mu}_{ij}\right)\psi_\mu,
\\
\dot\psi_i&=&-\delta \beta_i-\psi_{0,i}+2{\Gamma^\mu}_{i0}\psi_\mu, \\
\label{KN3}
\delta\g_{ij}&=&-2\psi_{(i,j)}+2{\Gamma^\mu}_{ij}\psi_\mu.
\end{eqnarray}
$\delta K_{ij}$ can be obtained by taking a time derivative of
(\ref{KN3}) and using the definition of the extrinsic curvature tensor
(\ref{ADM1}). These equations are coupled with algebraic, elliptic or
evolution equations for $\delta\alpha$ and $\delta\beta_i$ in terms of
$\delta\g_{ij}$ and $\delta K_{ij}$. Clearly the well-posedness of any
gauge defined in this way is a necessary condition for that gauge to
be well-posed together with any specific formulation of the Einstein
equations. 

\cite{KN} shows that the combination $\dot\alpha=\epsilon\dot K$,
$\beta^i=0$ (explicit parabolic $K$-driver with zero shift) is
ill-posed. By taking the limit $\epsilon=0$, it is also claimed
$K$-freezing lapse with zero shift is ill-posed, although this limit
is singular. Our results agree with these claims.  However, we only
consider the linearisation around Minkowski spacetime in Cartesian
coordinates, which simplifies the calculation, while in \cite{KN} all
lower-order terms in an explicit linearisation are kept, and are found
to give an unbounded ${\rm Re}\Lambda$ for the parabolic lapse driver
with fixed shift. On the other hand we find that $M(\omega_i)$ is not
diagonalisable when perturbing around the Minkowski background, while
diagonalisability is not checked in \cite{KN}.





\begin{thebibliography}{}

\bibitem{York} J. W. York, Jr, in {\it Sources of Gravitational
Radiation}, ed. L Smarr, Cambridge UP 1979.

\bibitem{NOR} G. Nagy, O. E. Ortiz and O. A, Reula, Phys. Rev. D {\bf 70},
  044012 (2004).

\bibitem{bssn2} C. Gundlach and J. M.  Mart\'\i n-Garc\'\i a, Phys. Rev. D
  {\bf 70}, 044032 (2004).

\bibitem{Yo2} H.-J. Yo, T. W. Baumgarte and S. L, Shapiro, Phys. Rev. D {\bf
66}, 084026 (2002).

\bibitem{HinderPhD} I. Hinder, PhD thesis, University of Southampton (2005). 

\bibitem{KST} L. E. Kidder, M. A. Scheel and S. A. Teukolsky,
  Phys. Rev. D {\bf 64}, 064017 (2001).

\bibitem{reduct} C. Gundlach and J. M. Mart\'\i n-Garc\'\i a,
  Hyperbolicity of second-order in space systems of evolution
  equations, e-print gr-qc/0506037, to be published in Class. Quant. Grav.

\bibitem{LindblomScheel} L. Lindblom and M. A. Scheel, Phys. Rev. D
  {\bf 67}, 124005 (2003).

\bibitem{SarbachBSSN} O. Sarbach {\it et al.}, Phys. Rev. D {\bf 67},
  064002 (2002).

\bibitem{SarbachTiglio} O. Sarbach and M. Tiglio, Phys. Rev. D {\bf
66}, 064023 (2002).

\bibitem{symmcoord} D. Garfinkle and C. Gundlach, 
Class. Quant. Grav. {\bf 16}, 4111 (1999).

\bibitem{Meudon} S. Bonazzola {\it et al.}, Phys. Rev. D {\bf 70},
  104007 (2004). 

\bibitem{SmarrYork} L. Smarr and J. W. York, Jr., Phys. Rev. D {\bf
17}, 2529 (1978).

\bibitem{Balakrishna} J. Balakrishna {\it et al.}, Class. Quant. Grav. {\bf
  13}, L135 (1996).

\bibitem{ReimannAlcubierre} B. Reimann {\it et al.}, Phys. Rev. D {\bf
  71}, 064021 (2005).

\bibitem{AEIcoordinates} M. Alcubierre {\it et al.}, Phys. Rev. D {\bf 67},
  084023 (2003). 

\bibitem{Shibata} M. Shibata, K. Taniguchi and K. Uryu, Phys. Rev. D {\bf
  68}, 084020 (2003).

\bibitem{Lindblompersonal} L. Lindblom, personal communication. 

\bibitem{Campanelli} M. Campanelli {\it et al.}, Phys. Rev. Lett. {\bf
  96} 111101 (2006).

\bibitem{Baker} J. G. Baker {\it et al.}, Phys. Rev. Lett. {\bf
  96} 111102 (2006).

\bibitem{Diener} P. Diener {\it et al.}, Phys. Rev. Lett. {\bf
  96} 121101 (2006).

\bibitem{Herrmann} F. Herrmann {\it et al.}, gr-qc/0601026.

\bibitem{BeyerSarbach} H. Beyer and O. Sarbach, Phys. Rev. D {\bf 70},
  104004 (2004).

\bibitem{BonaLehnerPalenzuela} C. Bona, L. Lehner and
  C. Palenzuela-Luque, Phys. Rev. D {\bf 72}, 104009 (2005).

\bibitem{Beyerlectures} H. R. Beyer, Beyond partial differential
  equations: A course on linear and quasi-linear abstract hyperbolic
  evolution equations, unpublished lecture notes (2005), gr-qc/0510097.

\bibitem{GKO} B. Gustafsson, H.-O. Kreiss and J, Oliger, {\it
Time-dependent problems and difference methods}, John Wiley, New York
1995.

\bibitem{AlcubierreBSSN} M. Alcubierre {\it et al.}, Phys. Rev. D {\bf 62},
  124011 (2000). 

\bibitem{KN} A. M. Khoklov and I. D. Novikov, Class. Quant. Grav. {\bf
  19}, 827 (2002). 

\bibitem{bssn1} C. Gundlach and J. M. Mart\'\i n-Garc\'\i a, Phys. Rev. D
  {\bf 70}, 044031 (2004).

\bibitem{thornexact} C. Gundlach, documentation for Cactus thorn
  ``Exact'', www.cactuscode.org (1998). 


\end{thebibliography}
\end{document}